%% file: ewsn-full.tex
\documentclass[10pt,emptycopyrightspace]{ewsn-proc}

\usepackage{graphicx}
\usepackage{balance}
\usepackage{comment}
\usepackage{epsfig}
\usepackage{xcolor}
\usepackage{amssymb}
\usepackage{amsmath}
\usepackage{xspace}
\usepackage{algorithm2e}
\usepackage{comment}
\usepackage{flushend}
\usepackage[hyphens]{url}
\usepackage[hidelinks]{hyperref}
\hypersetup{breaklinks=true}

\newcommand{\toolname}{AppStreamer\xspace}

\newcommand{\ie}{\emph{i.e.}\xspace}
\newcommand{\eg}{\emph{e.g.}\xspace}

\SetKwProg{Fn}{Function}{}{}
\date{}

\AtBeginDocument{%
  \providecommand\BibTeX{{%
    \normalfont B\kern-0.5em{\scshape i\kern-0.25em b}\kern-0.8em\TeX}}}

\numberofauthors{8}
\author{
\alignauthor Nawanol Theera-Ampornpunt \\
    \affaddr{Prince of Songkla University, Phuket Campus}\\
    \email{nawanol.t@phuket.psu.ac.th}
\alignauthor Shikhar Suryavansh \\
    \affaddr{Purdue University} \\
    \email{ssuryav@purdue.edu}
\alignauthor Sameer Manchanda \\
    \affaddr{University of Illinois at Urbana-Champaign} \\
    \email{schanda123@gmail.com}
\and
\alignauthor Rajesh Panta \\
    \affaddr{AT\&T Labs Research} \\
    \email{rpanta@research.att.com}
\alignauthor Kaustubh Joshi \\
    \affaddr{AT\&T Labs Research} \\
    \email{kaustubh@research.att.com}
\alignauthor Mostafa Ammar \\
    \affaddr{Georgia Institute of Technology} \\
    \email{ammar@cc.gatech.edu}
\and
\alignauthor Mung Chiang \\
    \affaddr{Purdue University} \\
    \email{chiang@purdue.edu}
\alignauthor Saurabh Bagchi \\
    \affaddr{Purdue University} \\
    \email{sbagchi@purdue.edu}
}

\title{{AppStreamer: Reducing Storage Requirements of Mobile Games through Predictive Streaming}}

\begin{document}
\pagestyle{plain}
\maketitle

\begin{abstract}
Storage has become a constrained resource on smart phones. Gaming is a popular activity on mobile devices and the explosive growth in the number of games coupled with their growing size 
contributes to the storage crunch. Even where storage is plentiful, it takes a long time to download and install a heavy app before it can be launched. This paper presents \toolname, a novel technique for reducing the storage requirements or startup delay of mobile games, and heavy mobile apps in general.
\toolname is based on the intuition that most apps do not need the entirety of its files (images, audio and video clips, etc.) at any one time. \toolname can therefore keep only a small part of the files on the device, akin to a ``cache'', and download the remainder from a cloud storage server or a nearby edge server when it predicts that the app will need them in the near future. \toolname continuously predicts file blocks for the near future as the user uses the app, and fetches them from the storage server before the user sees a stall due to missing resources. We implement \toolname at the Android file system layer. This ensures that the apps require no source code or modification, and the approach generalizes across apps.
We evaluate \toolname using two popular games: Dead Effect 2, a 3D first-person shooter, and Fire Emblem Heroes, a 2D turn-based strategy role-playing game. Through a user study, 75\% and 87\% of the users respectively find that \toolname provides the same quality of user experience as the baseline where all files are stored on the device. \toolname cuts down the storage requirement by 87\% for Dead Effect 2 and 86\% for Fire Emblem Heroes.
\end{abstract}

%
%

%

\input{sec_introduction}

\input{sec_related_work}

\vspace{10 pt}

\input{sec_design}

\vspace{10 pt}

\input{sec_components}

\input{sec_implementation}

\input{sec_experiments}

\input{sec_discussion}

\input{sec_conclusion}

%
%
\balance
\bibliographystyle{abbrv}
\bibliography{biblio}  
\end{document}

%% file: sec_introduction.tex
\section{Introduction}

The amount of content that users seem to want to store on their mobile devices has been outstripping the growth of storage available on the devices. For example, a survey conducted in 2017 found that a full quarter of the respondents said they deleted an app simply because their phone's storage space was full and they needed space~\cite{themanifest_2018}. 
In the developing world where less pricey smartphones are the norm, users often complain about having to uninstall applications or remove personal data. For example, the survey by SanDisk of smartphone users in India in 2018 found that 62 percent of smartphone users run out of storage space every three months~\cite{sandisk_2018}.

This growth is being spurred by two major factors. The first is the flowering of the app stores for both Android and iOS devices with a profusion of apps, and specifically, content-rich apps. One quantitative manifestation of this is that mobile accounts for approximately half of web traffic worldwide~\cite{mobiletraffic_statista_2018}.
One important class of heavy apps is mobile games, especially those with highly realistic graphics and videos. For example, at \$70.3 B, the mobile gaming industry accounts for more than half of all global gaming revenue in 2018 and mobile gaming apps accounted for 76\% of all app revenue~\cite{mobilegaming_mediakix_2019}.
These games take up large amounts of storage on our smartphones with Pokemon Go taking about 200 MB and more demanding games, like first person shooter (FPS) games taking 2-3 GB
~\cite{gamesize}.
The second factor is the increasingly powerful processors and graphics cards, high resolution displays, and ubiquitous network connectivity, which make it attractive to run these heavy apps and play these heavy games on mobile devices. 
Contrast the above state of affairs with the relatively slow growth in the amount of storage available on our smartphones. Well endowed smartphones today have 64--128 GB of internal storage, while the smartphones from 5 years ago had 16--32 GB standard, with a big chunk of this used by the OS itself and built-in apps that cannot be removed.
Proposed replacements such as 3D NAND technology are open to the vagaries of technology development and adoption.


{\em To sum up, the trends indicate that our desire to store content on smartphones is going to continue to outstrip the amount of inbuilt storage capacity available on these devices.}

\noindent {\bf Available solutions to storage crunch} \\
There are two consumer-grade options available today for alleviating the storage crunch. The first solution approach is the expansion to microSD cards through an optional expansion slot on the device.
However, this is not always possible (\eg, iPhones and Google Pixel do not have microSD card slot), incurs extra cost, and there are restrictions on putting apps on SD card. 
Original manufacturers often prevent their built-in apps and core functionality from being moved to the expandable storage~\cite{marshmallow_microsd}, as do many game vendors~\cite{androidpit-internal-storage-vs-microsd}, for copyright or performance reasons (internal storage read speed can be as high as 11X speed of SD card read)~\cite{microsd, microsd_speed}. 
The second solution approach is the use of cloud storage. The most popular use case is the storage of photos and static content on cloud storage. For example, the Google Pixel phone has a widely advertised feature that its photos will be backed up to Google Cloud, at full resolution, with no charge to the end user. This second solution approach however is less applicable for mobile applications. If an application's files and data are stored on the cloud rather than on the phone, when the user wants to run the application, the entire application must be downloaded before it can be used. This adds unacceptable amount of delay to application startup (18 seconds for an average 40 MB application, based on average worldwide LTE speed).

An alternative approach is running the application on a cloud server, and streaming the processed contents to the phone to display on its screen. This approach is widely in use today for gaming and is referred to as {\em cloud gaming}~\cite{huang2014gaminganywhere, outatime}. Expectedly, it has high bandwidth overhead, and incurs input delay due to network latency and video encoding/decoding overhead. These factors make this a challenging fit for highly interactive applications such as games, especially on wireless networks (WiFi or cellular) where bandwidth and latencies can be quite variable. We quantify this overhead through experiments on our two evaluation games run on a cloud gaming platform in Section \ref{sec:prior-comparison}.


\noindent {\bf Our Solution Approach: \toolname} \\
In this paper, we introduce the design of \toolname, which alleviates the storage crunch on mobile devices for larger-sized applications. It does this through extending the traditional microarchitectural notions of prefetching and caching from local storage to cloud storage. 
In our context, the device's local storage is analogous to the cache, while the cloud storage is the farther-off storage which has high access latency but contains all data. In contrast to prefetching techniques used by modern operating systems, \toolname's prefetching has to be much more aggressive as cache misses can easily ruin the user experience by introducing noticeable delays. For example, delays of a few tens of milliseconds are considered intolerable for the gamer \cite{card1991information}.
Even when local storage is not a constrained resource, \toolname can reduce the startup cost of an app when it is being executed for the first time (a common enough occurrence as surveys point out~\cite{pcquest_smartphone_usage,themanifest_2018})
Even if a user chooses cloud gaming, \toolname can 
mitigate stalls by predicting and pre-fetching the required blocks. 

\begin{figure}[tb]
\centering
\includegraphics[clip,trim=0cm 6cm 2.2cm 0cm,width=\columnwidth]{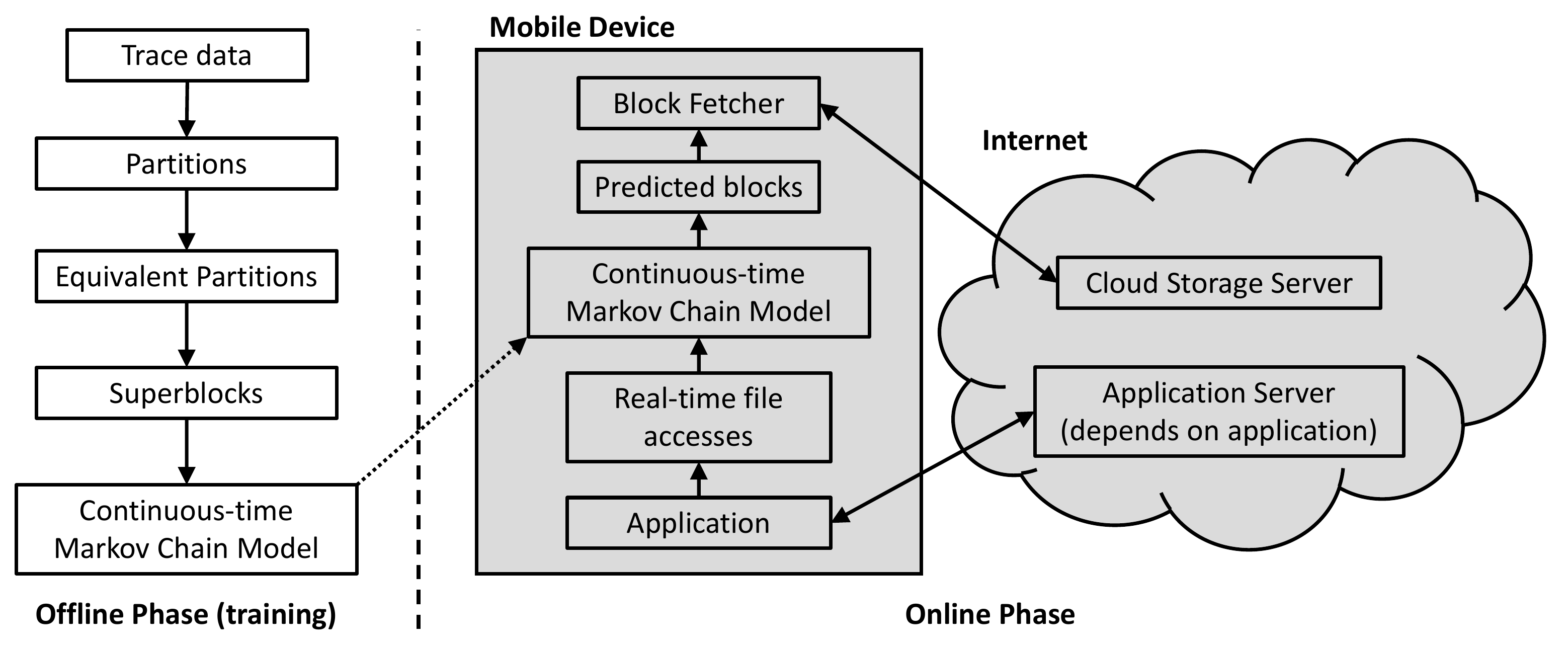}
\vspace{-1pt}
\caption{Overview diagram of \toolname, showing the components on the mobile device and the cloud, as well as the offline training and online phases.}
\vspace{-13pt}
\label{fig:overview}
\end{figure}

\toolname predicts the use of data blocks in the near future, based on file usage patterns of the application. The predicted blocks are then prefetched from the cloud storage with the goal of ensuring each data block is present on the device's storage {\em before} it is needed by the application. A high level overview of \toolname is shown in Figure~{\ref{fig:overview}}, which depicts the client mobile device and the server infrastructure, namely the cloud server which stores the application parts to be streamed to the device and optionally, an application server in the cloud (that we do not modify). In the figure, we also show the offline training of our model and the online processes of predicting and pre-fetching the required parts of the application. 

\begin{figure}[tb]
\centering
\includegraphics[clip,trim=3.7cm 17.7cm 5cm 3.6cm,width=0.9\linewidth]{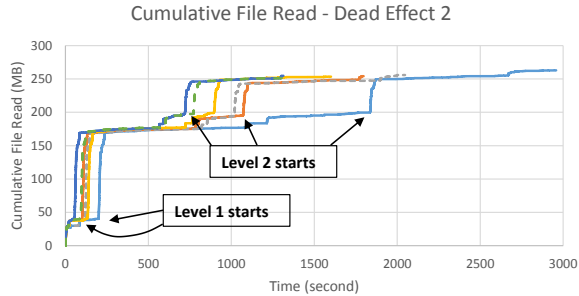}
\caption{Cumulative file read in six playthroughs of the first two levels of Dead Effect 2}
\label{fig:dead_effect_2_runs}
\end{figure}

We develop the concepts in the context of interactive games, because they are typically large in size and require high interactivity and thus low latency, though the concepts as well as the implementation also apply to other types of applications on mobile devices\footnote{Note that a streaming media application, like YouTube client, does not fall within this scope because in it, the storage-intensive part is the content, which is streamed to memory and removed when the session ends, and thus does not consume storage on the device.
}.
Empirically, we find that file blocks are accessed in granularities of tens of MB for most of the interactive games, 
as can be seen for the game Dead Effect 2 in Figure \ref{fig:dead_effect_2_runs}. 
The jumps in the curves correspond to chunks of application content that are needed for the game, under different runs. Hence, \toolname has to perform accurate prediction of which chunks will be needed and initiate the prefetching with significant lookahead, considering that typical cellular network bandwidth is low, in the range 10-20 Mbps---the US average for LTE bandwidth is 13.95 Mbps~\cite{opensignal}.

Our solution works in a manner that is agnostic to the actual game and thus needs no semantic understanding of the game or access to the source code of the game, either on the client or the server side.
The input parameters to \toolname allow for an exploration of the two dimensions of the tradeoff---the amount of storage used on the device and the bandwidth used to bring data to the device versus the pause time inflicted on the user. 

{\em We posit that \toolname presages the model of streaming applications, paralleling the current ubiquitous trend of streaming content for mobile devices.} An analogy, though imperfect, can be drawn to the change that happened in the media streaming field where in the late 1990's/early 2000's, there was a move toward streaming audio (remember Real Networks) and then streaming video. At that time, the practice had been to download entire media first before listening to it/viewing it (\eg, iTunes in its early incarnations only supported downloaded media). We hypothesize that similarly we will have a trend in certain niches (such as, heavy duty users) to stream required resources for the applications when needed. This is because for games as well as for applications, most users only use a small fraction of the available scenes (games) or functionality (applications) at any one time.  

\noindent {\bf Evaluation} \\
We evaluate the effectiveness of \toolname through two distinct types of games customized for mobile devices---Dead Effect 2, a FPS game and Fire Emblem Heroes, a turn-based strategy role-playing game (RPG). We conduct user studies for both of these games and find that in the majority of the cases, the users do not find any significant degradation in quality of experience. 
This is achieved while using 86--87\% 
less storage than the baseline. We also conduct microbenchmark studies to evaluate the effect of various configuration parameters on the two dimensions of the tradeoff, which guides choice of the values for the parameters.

In summary, we describe the following contributions in this paper.
\begin{enumerate}
\addtolength\partopsep{-1em}
\addtolength\topsep{-1em}
\addtolength\leftmargin{-1em}
\item We create a middleware for mobile devices called \toolname that reduces the amount of download needed during installation as well as local storage space needed while executing an application.

\item We show that it is possible to record data block accesses and use them to predict what blocks will be needed in the future and how far ahead in the future. We achieve this in a way that is generalizable across players and across games and does not require source code access.

\item We evaluate the effectiveness of \toolname through two popular, closed-source games and microbenchmarks. We find that \toolname is effective in reducing the storage requirement on the device, while not degrading the user quality significantly for this highly latency-sensitive workload.
  
\end{enumerate}

%% file: sec_related_work.tex
\section{Related Work}
\label{sec:related}

In this section we will discuss current solution approaches for mobile gaming as well as related pieces of work.

Our work is complementary to two closely related classes of work. 
One class of work \cite{outatime,cuervo2015kahawai} makes modifications to the game at both the client and the server side and collaboratively gets the right frame to the device at the right time.
The second class of work \cite{gordon2012comet, shi2014cosmos} offloads computationally demanding parts of the local application to a cloud server.
\toolname can use solutions from the first class to assist in the pre-caching decisions at the device, by using the resources at the server. Thus the computational load at the client can be reduced and close to optimal lookahead can be used for pre-caching content. \toolname can leverage solutions from the second class to offload some demanding parts of the game computation, such as, computing the rendering of some graphic-rich frame, to a nearby cloud server. Since we predict some blocks that will be used in the near future, this can help in making the appropriate offloading decision at the appropriate time.

One approach to mobile gaming is the thin client model of gaming, often called ``Cloud Gaming''~\cite{huang2014gaminganywhere}.  In this model, the mobile application transmits user inputs to a server, which performs the computations and graphics rendering. The rendered frames are then streamed to the mobile device as a video. This approach has the advantage of removing the computational as well as storage limitations of mobile hardware. However, the bandwidth usage is high and delays in interactivity due to latency can make this approach unappealing~\cite{chen2014cloud}, as 100--200 ms of round-trip time is common in mobile networks. We perform experiments to compare \toolname with traditional cloud gaming approach and found that our approach has 77\% lower bandwidth usage and 98\% lower latency.

Lee {\em et al.} proposed reducing cloud gaming latency by using Markov modeling and Kalman filtering to speculate about future frames \cite{outatime}; however, this approach requires modifying the source code of the game, which can be unappealing to developers, and may not be an option for end users since most popular games are closed source.
Abe {\em et al.} propose the idea of streaming virtual appliance (VA) on mobile network by caching files predicted to be necessary in the near future~\cite{abe2013vtube}. The overall idea is similar to \toolname, however, we focus on highly interactive mobile applications as opposed to all applications in general. Our goal is to make the user experience indistinguishable from storing all files locally while their approach results in long, 1--3 minutes start-up delay, and multiple interruptions during each session.
Further, their approach applies to Linux clients (ours is for Android devices), relies also on temporal locality across user sessions (for us only locality within a session is useful), and the predictability of file access is higher (and thus the problem is simpler).

We take inspiration from the broad base of work on file prefetching since at a high level, the problem of efficient caching is the same as in \toolname, except that we are working at a different layer in the memory hierarchy. That is, we use the device's local storage as the cache for the cloud storage. The first approach to file prefetching, application-directed prefetching requires non-portable special directives from the application developer, and thus cannot be applied to existing applications~\cite{mowry1991tolerating}.
In many cases, the developer needs to predict user behavior for such systems to work well, and this can be difficult. History-based approaches use file access history to predict file accesses. These works share a similar workflow with \toolname, but are either very limited in scope (SuperFetch optimizes application launch, and Hot Files feature in HFS Plus moves small frequently-used files closer to metadata)~\cite{russinovich2007inside, harter2012file}, work at too coarse a granularity (whole file instead of block)~\cite{tait1991detection, griffioen1994reducing}, have unrealistic assumptions (between any two user requests, there is enough time to prefetch as many pages as wanted), or require extensive user input (Coda, the distributed file system, required user listing of all files required offline)~\cite{kistler1992disconnected} and were not evaluated in real systems~\cite{vitter1996optimal}. Thus, \toolname has a different context and requirements and does not have as strong assumptions.

\vspace{-1em}

%% file: sec_design.tex
\section{Design Overview}
\label{sec:design}

The goal of \toolname is to reduce storage requirements of large mobile applications. This is done by storing file blocks that are needed by the application on the mobile device, and speculatively fetching more blocks from a cloud storage server as the application runs.  
We focus on files that are part of the application package. This includes required files that some applications (especially games) download on the first run. The key characteristic of these files is that they are read-only and are needed across all users and across different executions. Files written by the application are small and contain user-specific contents which cannot be shared across users, so we always store these files on the device.

We would like the system to work for all applications without any modifications or access to the source code. Thus, \toolname is  implemented at the operating system level. As long as the blocks the application reads are already on the device, the application will operate normally without any user-perceptible stall, even when the entirety of the application blocks is not present on the device. Intuitively, past file reads should be a good predictor of future file reads, so this is what we use for our prediction model. It is possible to include other sources of the application's behavior from the perspective of the operating system, such as CPU usage, memory usage, network usage, and past file writes, 
but we have found that past file reads alone already provide good prediction accuracy. There is also a desire to minimize the monitoring overhead which contributes to possible slowdown of the application. 

\toolname's operation can be divided into two phases: offline phase and online phase, as shown in Figure~{\ref{fig:overview}}. The figure shows the cloud storage server, which is a component introduced due to \toolname. The server is expected to have {\em all} of the content needed by the application, prior to the execution of the application on the mobile device. This can also be an edge server or a hybrid as proposed in~\cite{suryavansh2019tango}. In the offline phase, the file access prediction model (Continuous-time Markov Chain in this case) is trained using a collection of file access traces from multiple users using a specific application. Note this does not need to include the specific user who will use the device in the online phase. As long as the runtime path is a combination of one or more paths from training, the model will be able to combine the patterns so that it has knowledge about all paths. In the online phase, as the user opens up and uses the application, the model predicts blocks that are needed in the near future, and fetches them from the cloud storage server in real time. 


While \toolname is agnostic to the type of application whose storage requirement is to be reduced, most of the large mobile applications today are games, due to their rich media content and large number of control paths. Therefore, from this point onward, we will focus on mobile games running on smartphones and tablets as our target application domain. 


\vspace{-1em}

%% file: sec_components.tex
\section{Components of \toolname}
\label{sec:components}

\toolname is composed of three main components: 
a component for capturing file blocks accessed by the application, file access prediction model, and data block fetcher that downloads data blocks from the cloud storage server. In this section, we describe the design of each component and how they interact among themselves.

\subsection{File Access Capture}
\toolname reduces storage requirement of applications by storing the file blocks that will be read by the application on the mobile device. 
As long as we can predict the application's file reads and fetch the required blocks before a request from the application, the application will work normally as if all files are present on the device. We use past file reads to predict an application's file accesses.
To capture the file read information, we create a wrapper function for the file read function at the virtual file system layer of the Android operating system. Information captured includes file path, file name, file pointer position, number of bytes to be read, and timestamp. From the file pointer position and the number of bytes read, we can determine which blocks are being read. 
In the offline (training) phase, the file reads are logged and written to a file, which will be used to train the prediction model. In the online phase, the file reads are given to the prediction model as input, in order to make predictions in real time.

\subsection{File Access Prediction Model}
The file access prediction model is trained using file access traces collected from multiple users playing a specific game. The raw input to the prediction model is the file read system calls the game makes. The input features associated with each call are file path, file name, file pointer position, number of bytes to be read, and timestamp.
Because Android's file system operates at the block level, where each block is 4 KB, we convert each record of file read system call into individual 4 KB blocks.

A model is needed to capture the file access patterns as well as predict future file accesses. 
We believe an appropriate model for these challenges is Continuous-Time Markov Chain (CTMC). CTMC captures the possible transitions between states as well as the duration spent at each state in a probabilistic manner. The duration information allows us to limit the lookahead when making predictions
Markov models are probabilistic in nature and the predictions give us the probability associated with each predicted block. This gives us the ability to adjust the tradeoff between delay (due to a necessary block not being present) and potential waste of bandwidth and energy. The computational and memory overhead, as well as the accuracy of the prediction, depend on what we define as the states for the model.

The straightforward way to define the state of the CTMC is to use individual blocks read as the states. However, the memory, storage, and computational requirements become too high. Consider that a 1 GB game has 250,000 4 KB blocks. In the worst case, we need to store close to $250{,}000^2=62.5$ Billion transition probabilities. Making predictions with the CTMC requires a depth-first-search, and having a branching factor of 250,000 makes the computation infeasible even at low depth limit, and the necessary depth limit can be quite high in this case. 
We empirically show that a model that uses individual blocks as states is infeasible in Section \ref{sec:prior-comparison}.

Instead of using individual blocks read as the states, we use groups of blocks as the states. From the file access traces in Figure \ref{fig:dead_effect_2_runs}, we notice that data blocks tend to be accessed in groups, and these groups are similar across different users and different runs. Therefore, we design an algorithm for grouping blocks together, as well as extracting common patterns from different runs. We find that this significantly reduces the memory, storage, and computational requirements of the model.


\subsection{Block Grouping}
As mentioned above, grouping blocks is an important step for state space reduction for the CTMC. 
A group of blocks is simply blocks that are frequently accessed together in multiple different runs. We want each group to be as large as possible (to reduce the state space), while keeping the property that most blocks in the group are accessed together for at least a few runs. Our specific algorithm, shown in Algorithm \ref{algo:buildbeta} creates such groups when the number of blocks in a group times the number of runs accessing all the blocks in the group is sufficiently large. 
Note that there can be overlap between groups. We propose a method for grouping blocks that meets these requirements and essentially summarizes the raw block accesses into essential information that the CTMC can easily process.

\begin{figure}[tb]
\centering
\includegraphics[width=0.95\columnwidth]{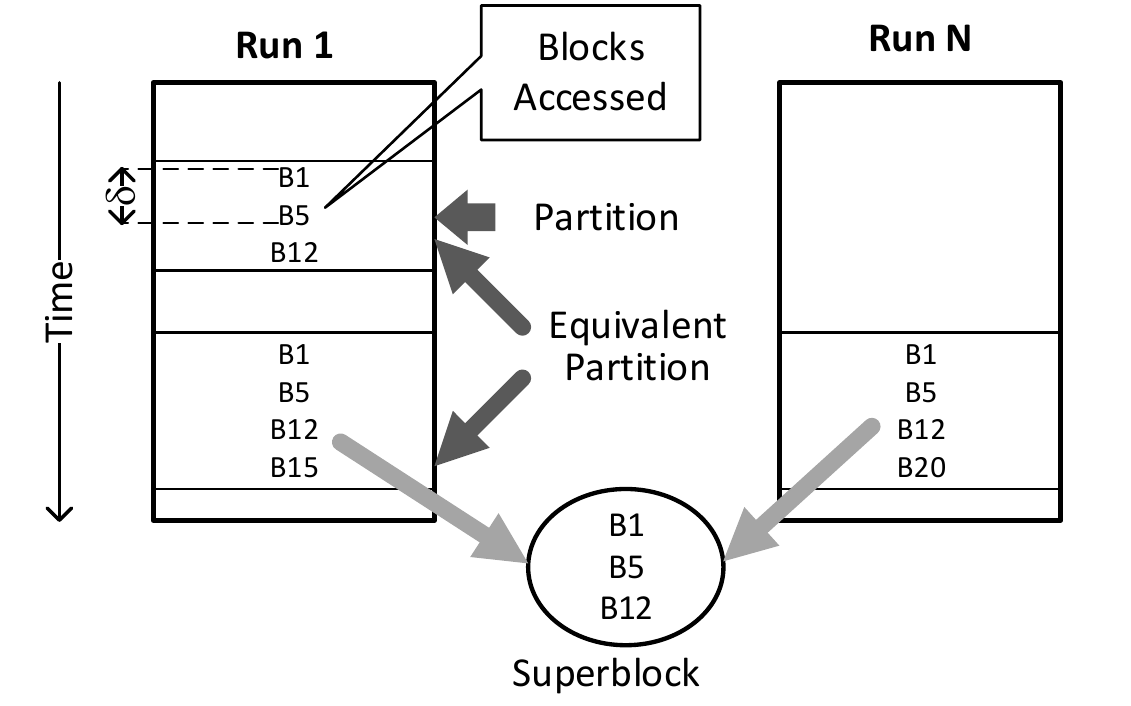}
\caption{Grouping of blocks during training. Concepts of {\em Partition} and {\em Equivalent Partition} within a single trace and {\em Superblock} across multiple traces.}
\label{fig:block-grouping}
\end{figure}

Our block grouping method can be divided into three steps as shown in Figure \ref{fig:block-grouping}. In the first step, we group block reads in one trace that are close together in time into {\em partitions}. In the second step, we generate \emph{equivalent partitions} from the partitions by merging partitions that are sufficiently similar. In the third step, we extract overlaps in equivalent partitions from different trace runs to create groups of blocks which we call {\em superblocks}. Superblocks represent groups of blocks that are frequently accessed together, and they are used as input to the CTMC.

\subsubsection{Partitions}
In this step, each trace run in the training data is processed separately to produce \emph{partitions}. Each partition contains blocks that are accessed close together in time, with the threshold of maximum time between successive block reads specified by a parameter $\delta$.

\subsubsection{Equivalent partitions}
Empirically, we found that some partitions within the same trace run are very similar to each other. In order to remove redundant information, we merge near-identical partitions from the same trace into one {\em equivalent partition}. We use Jaccard index as the similarity metric with each partition considered as a set of data blocks. 
A Jaccard index of 1 means that both sets are identical, while a Jaccard index of 0 means there is no overlap between both sets. All pairs of partitions with Jaccard index higher or equal to a parameter denoted by $\tau$ are merged into one. No changes are made to partitions that are not sufficiently similar to any other partitions. 
After these steps, we have the equivalent partitions, which are used as input to the superblock step.
The value for the parameter $\delta$ should be chosen to separate block reads in different ``batches.'' Empirically, the value should be between 1--1000 milliseconds. We discuss choosing $\delta$ and $\tau$ further in the microbenchmark section (Section \ref{sec:microbenchmarks}).

\subsubsection{Superblock}
Equivalent partitions generated in the previous step represents groups of blocks that are frequently accessed together within a single run.
At this point, we essentially have one model per run. In order to utilize information about different playing styles and different execution paths, we need to combine information from different runs into a single model, while removing redundant information. Due to different timings and execution paths, equivalent partitions in multiple trace runs are not exactly identical. For example, an equivalent partition in one trace run can be made up of three equivalent partitions from another trace run, or may not show up at all in yet another trace run. Intuitively, we want to find large intersections of equivalent partitions across trace runs, consider the intersection as a single unit of blocks, and remove the intersection from the equivalent partitions that contribute to the intersection. We then repeat the process until no large intersection can be found. 
We call each large intersection a {\em superblock}. The challenge comes from the fact that the blocks in some equivalent partitions may not be present in all trace runs.

The pseudocode for creating superblocks is shown in Algorithm~\ref{algo:buildbeta}. First, we find the largest overlap between any $n$ equivalent partitions where each belongs to a different trace and $n$ can be lower than the number of traces. The intuition behind not requiring $n$ to be exactly the number of traces is that the playing style, and thus the file access patterns can be very different across traces, so a certain common pattern may not be present on all traces. For the purpose of this algorithm, the size of an overlap is defined as the number of blocks in the overlap multiplied by the number of contributing traces $n$. 
The blocks in this overlap are then removed from \emph{all} of the contributing equivalent partitions, and put aside as a new superblock. The process repeats until the size of the overlap is lower than a threshold $minSuperblockSize$. At this point, the remaining equivalent partitions are merged into the superblocks that are closest in time. Recall that a superblock is the intersection of multiple equivalent partitions from different runs. For each contributing equivalent partition, the superblock takes the timestamp from it. For example, a superblock that comes from three runs will have three timestamps, each corresponding to a run. To determine which superblocks are closest in time to each remaining equivalent partition, the timestamp corresponding to the equivalent partition's run is used. 
At the end, each trace is transformed into a sequence of superblocks, which are then used as input to the Markov model.

{
\begin{algorithm}[t]
\footnotesize
\LinesNotNumbered
\Fn{createSuperBlocks(equivPartitions, numTraces, minSuperblockSize)}{
superBlocks $\leftarrow$ [ ]\;
\While{true}{
  overlap $\leftarrow$ findLargestOverlap(equivPartitions, numTraces, 1, null)\;
  \If{overlap.size $<$ minSuperblockSize}
  {
    break\;
  }
  remove blocks in \textit{overlap} from the equivalent partitions it originated from\;
  superBlocks.append(overlap)\;
}
\tcc{Merge each remaining equivalent partitions into the closest superblock}
\For{i $\leftarrow$ 1 to numTraces}
{
  \For{j $\leftarrow$ 1 to length(equivPartitions[i])}
  {
    \If{equivPartitions[i][j].size $>$ 0}
    {
      merge equivPartitions[i][j] into the closest superblock, based on timestamp of original equivalent partition from trace $i$\;
    }
  }
}
\Return superBlocks\;
}
\caption{Create superblocks from equivalent partitions}
\label{algo:buildbeta}
\end{algorithm}
}

\subsection{Markov Chain}
A Markov chain is specified by a number of states, the transitions among the states, and initial observation distribution. To capture the time duration spent in a state, we add duration which is associated with each possible state transition to the model. However, unlike the usual CTMC where the duration follows an exponential distribution, we use the arithmetic mean and standard deviation of the duration seen from the training data. Each superblock corresponds to a state in the CTMC. Time spent in a duration is independent of other parameters of Markov chain, which can be learned through the usual method. Although our model's formulation is slightly different from the original CTMC, for simplicity we will continue to refer to our model as CTMC.

To determine the current state, recent file reads are partitioned in the same way partitioning is done for training. With the most recent partition, we find the superblock(s) that are a subset of this partition. If there are multiple superblocks, any arbitrary choice works well in practice because the lookahead is long enough that small differences in the starting state do not lead to widely divergent paths. The chosen superblock becomes the current state of the CTMC. In order to make predictions, from the current state, we perform a depth-first search to compute the probability of each possible following sequence of future states. However, the number of possible sequences grows exponentially with the length of the sequence. To keep this computation feasible, we prune the search in two ways: by stopping when the probability of reaching that particular state through a specific path is lower than a threshold $p_{stop}$
or when the cumulative duration exceeds the lookahead time $L$. At the end, we have the probability of each superblock being read within the lookahead time. If this probability is higher than the block fetching threshold, denoted by $p_{download}$, then the blocks in the superblock that are not already on the device are added to the download queue.
This parameter controls the tradeoff between downloading unnecessary blocks and the likelihood and amount of delay experienced by the user.

Since gamers with different skill levels often play games at different speeds, while the sequence of file block accesses may be the same, we take into account the user's playing speed when making prediction. As the user plays the game, we observe time duration spent at each state, and compare it to the average duration learned by the model. Because playing speed only affects the duration between state transition, the effect of shorter duration can be simulated by increasing the lookahead time $L$. Therefore, when we make predictions, the lookahead time is adjusted, with a faster player getting higher lookahead, proportional to how fast she plays compared to the average.

Similarly, different players may play in different ways, and this can result in different control paths being taken. Such information can also be captured by a {\em single} Markov model, by learning the probability of transitioning to state $j$ given that the current state is $i$ for all pairs of $(i, j)$. 
This lets us utilize a single model for everyone,
instead of having multiple models for each class of users, which would then require a classifier that determines the class of current user, as well as more training data. 

\subsection{Data Block Fetcher}
The data block fetcher is responsible for downloading data blocks from the cloud storage server. There are two types of data blocks that need to be downloaded---predicted data blocks for future needs and application-requested data blocks for current needs. The predicted blocks are stored in a download queue in first-in, first-out (FIFO) order. The speculative block fetcher runs as a kernel thread and downloads predicted blocks in the download queue, using lowest network priority in order to not affect the responsiveness of other processes. 
Application-requested blocks need to be fetched urgently because the application, and by extension the user, is waiting for its response and so a request is sent immediately to the cloud storage server. The original read system call will block until the data is obtained. 

\subsection{Initial Files Cached}
On Android, an application's APK (Android Package Kit) is always immediately read in its entirety when the application is launched, so we always store the entire APK on the device. Some other resource files may also be accessed at launch, or soon after. Since we do not have enough time to fetch these files during the short duration between application launch and when they are accessed, they must also be stored on the device. We denote the amount of file blocks that are stored on the device at all times (excluding the APK) by $B_{initial}$. This parameter controls the tradeoff between delay and amount of storage used. 

\subsection{Temporary Storage Limit}
In the current state of practice, {\em all} of a game's files are {\em always} stored on the device. With \toolname, only the APK and a small portion of the game's files 
(specified by $B_{initial}$) 
are always stored on the device. These files that are always stored on the device take up {\em permanent storage} (until the game is uninstalled). With \toolname, predicted blocks that are fetched speculatively are also stored on the device. However, they are deleted when the game's process terminates. Thus, we can say that these fetched blocks take up {\em temporary storage} on the device. As long as the user does not play multiple games at the same time (which is uncommon), this temporary storage can be shared among all games that uses \toolname. 

In the case where available free space is extremely limited, a hard limit on the temporary storage is imposed. In this case, when the limit is reached, 
we evict old blocks using the least recently used (LRU) eviction policy. Blocks that are predicted (by \toolname) to be read in the near future are never evicted. This hard limit can have an effect on bandwidth usage and may introduce extra delays compared to when there is no limit. Such effects are quantified in our microbenchmarks (Figure \ref{fig:micro1} ``Temporary storage limit'').

%% file: sec_implementation.tex
\section{Implementation}
In this section, we describe how different components of \toolname are implemented in the Android system. Although different OSes may use different file systems, \toolname only requires that files be organized in blocks, so it can be implemented in most OSes.

\subsection{Capturing File Accesses}
File read is the input to \toolname's file access prediction model. To collect the file reads, used in both offline training and online prediction phases, 
we create a wrapper function for the file read system call \texttt{vfs\_read} at the virtual file system layer of the Android kernel. In the offline phase, the wrapper function records the file read if it is a relevant file (\ie, a file used by that game that is common across users), 
in a log file, before doing the actual file read and returning the requested data as usual. Information recorded includes file name and path, current file pointer, number of bytes requested, and timestamp. 
Empirically, we found that the overhead of logging is not noticeable to the user.
From our experiments, the average size of uncompressed log is 2.93 MB for 30 minutes of gameplay. It can be easily compressed if necessary due to repeated path and file names.
In the online phase, the wrapper function passes on the information about the file read to the prediction model. If the requested blocks are already on the device, they are returned immediately to the caller. If the requested blocks are not on the device (\ie, a ``cache miss''), the urgent block fetcher immediately requests the data from the cloud storage server. 

\subsection{Block Storage}
Most file systems organize files as a collection of blocks. It is therefore natural for \toolname to also operate at the granularity of blocks. In Android, each block is 4 KB in size. To simplify organization of blocks, we first create a file which acts as a container for all blocks and \toolname creates a mapping from the game's accessed file name and block offset to the location in this file.
A hash table in memory stores the mapping and since a game uses a limited number of blocks in its working set, the size of this hash table is small--with a 1 GB game and a liberal 25\% of its blocks in the working set, the hash table is only 1 MB in size.

%% file: sec_experiments.tex
\section{Experimental evaluation}
\label{sec:evaluation}

We evaluate \toolname in two ways: with user studies and with microbenchmarks. 
The goal of the user studies is to evaluate \toolname in a realistic setting, and measure the effect of delays introduced by \toolname by having users rate the user experience. 
The goal of the microbenchmarks is to evaluate how different parameters of the model affect the overall performance of \toolname. Due to the large number of parameter values explored, the microbenchmarks are done using trace-based simulation.

\subsection{Games Used for Evaluation}
We use two Android games in our evaluation: Dead Effect 2 and Fire Emblem Heroes. Dead Effect 2 is a 3D single-player first-person shooter game. Gameplay is divided into multiple levels, where the blocks needed for each level are loaded at the beginning of the level. The levels are linear (level 1, 2, 3, and so on), but different collected traces show diversity among users of the blocks accessed during gameplay. Its APK is 22.91 MB and all of its resources are stored in a single OBB (opaque binary blob) file which is 1.09 GB.

Fire Emblem Heroes is a 2D strategy role-playing game. Gameplay is divided into multiple small levels in several game modes. At first, only the main story mode can be played. As the player completes a few levels, paralogue mode, training mode, special maps mode, and player-versus-player mode are unlocked. These modes can be switched to and from at any time and in any order. 
The players choosing levels affects which blocks are needed, and also makes prediction in Fire Emblem Heroes nontrivial. The game has roughly 5,200 data files, whose sizes sum up to 577 MB, including the 41.01 MB APK.
We chose these two games as they represent two dominant classes of games on mobile devices, with differing characteristics in terms of themes, player interaction, and control flow. Both have the common characteristics of heavy graphics and low latency interactions.

\subsection{Training Data}
For Dead Effect 2, the trace data for the user study consists of 6 runs collected from two players. For the microbenchmarks, the trace data consists of 12 runs collected from four players. Each run is from start of the game to the end of level 2, which takes roughly 30 minutes.

The file read pattern of Dead Effect 2 is shown in Figure~\ref{fig:dead_effect_2_runs}. As soon as the game is launched, the entire APK is read and a small part of the OBB file is read. When each level starts, the resources needed for that level are loaded at the beginning, resulting in a jump in the cumulative file read. During each level, there are several small jumps dispersed without any easily perceptible pattern.
For Fire Emblem Heroes, the trace data for the user study consists of 7 runs collected from one player. For the microbenchmarks, the trace data consists of 14 runs collected from four players. For this game, each run consists of 20 minutes of gameplay from the beginning where the player is free to choose which level to play.


\subsection{User Study - Dead Effect 2}
Performance of \toolname depends on the network speed. For the user study, we set up the storage server on a local machine with network speed limited to 17.4 Mbps, which is the average worldwide LTE speed reported by Open Signal as of November 2016~\cite{opensignal}. The phones used are Nexus 6P running Android 6.0.1. Each participant plays the first two levels of the game, once on a phone with \toolname, and once on an unmodified phone. Each user then filled a questionnaire with four questions: 1) user's skill level (in that category of games, such as FPS), 2) quality of user experience, 3) delays during menu and loading screens, and 4) delays during gameplay inside a level. 

\begin{figure}[tb]
\centering
\includegraphics[clip,trim=2cm 9cm 1.7cm 9cm,width=0.8\columnwidth]{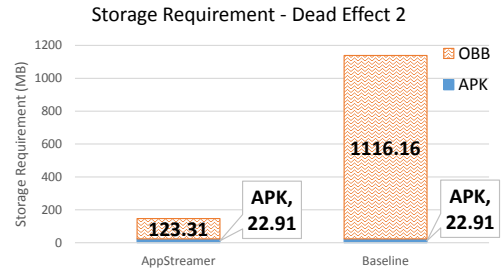}
\caption{Storage requirements for Dead Effect 2}
\label{fig:storage1}
\end{figure}

The first user study is done with Dead Effect 2, with 23 users participating in the study. The amount of storage used by Dead Effect 2 is shown in Figure~\ref{fig:storage1}. Baseline refers to the current state of practice which is storing the whole game on the phone. The amount shown for \toolname corresponds to the permanent storage used by files that are always stored on the phone. As the user plays the game, more blocks are downloaded on the fly. These blocks are stored in the temporary space and not shown in the figure. Overall, \toolname uses 146.22 MB of permanent storage, compared to 1,139.07 MB for baseline. This represents a 87\% storage saving.

\begin{figure*}[tb]
\centering
\includegraphics[clip,trim=8cm 10.2cm 4cm 10.2cm,width=0.4\columnwidth]{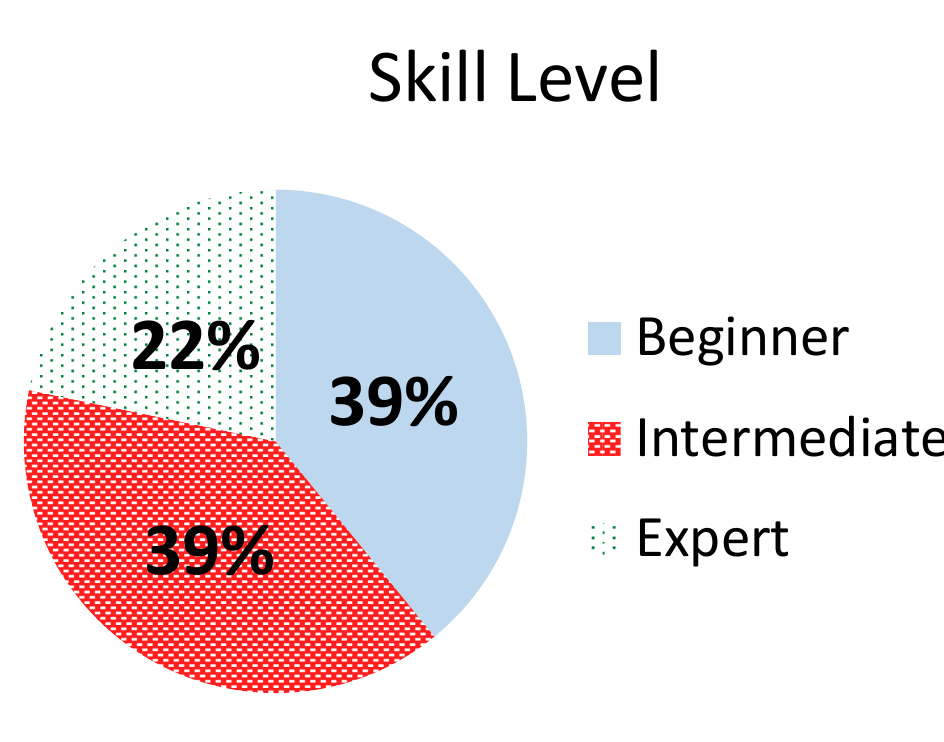}
\includegraphics[clip,trim=8cm 10.4cm 2cm 10.2cm,width=0.5\columnwidth]{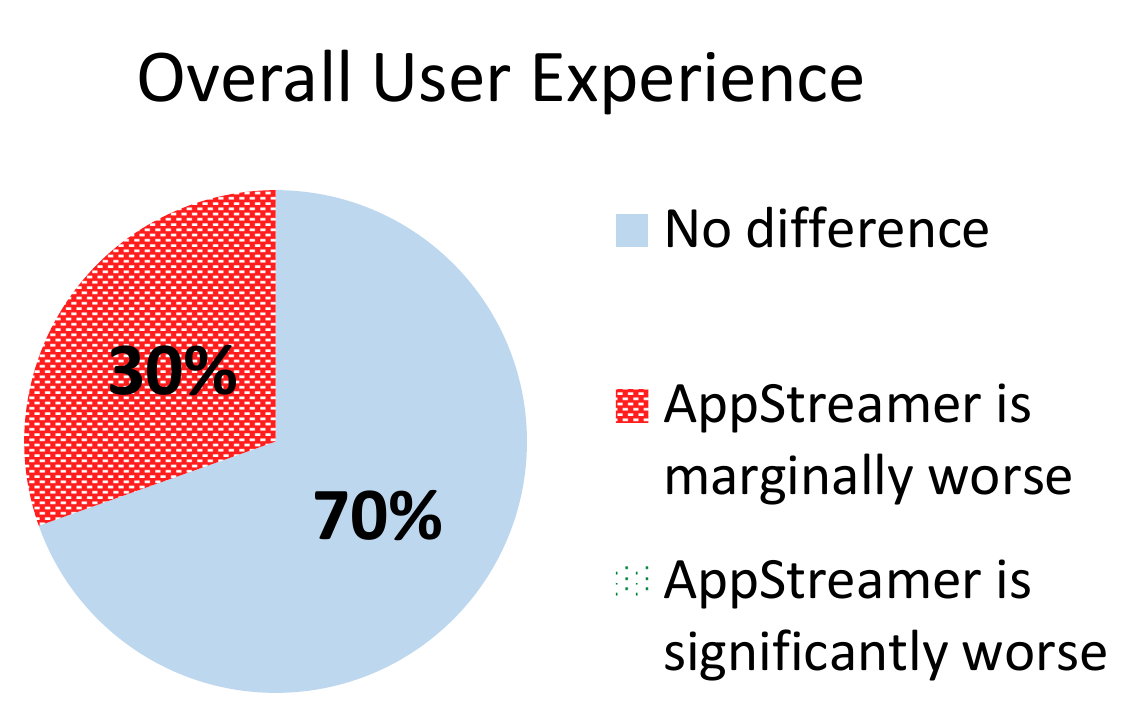}
\includegraphics[clip,trim=8cm 10.4cm 2cm 10.2cm,width=0.5\columnwidth]{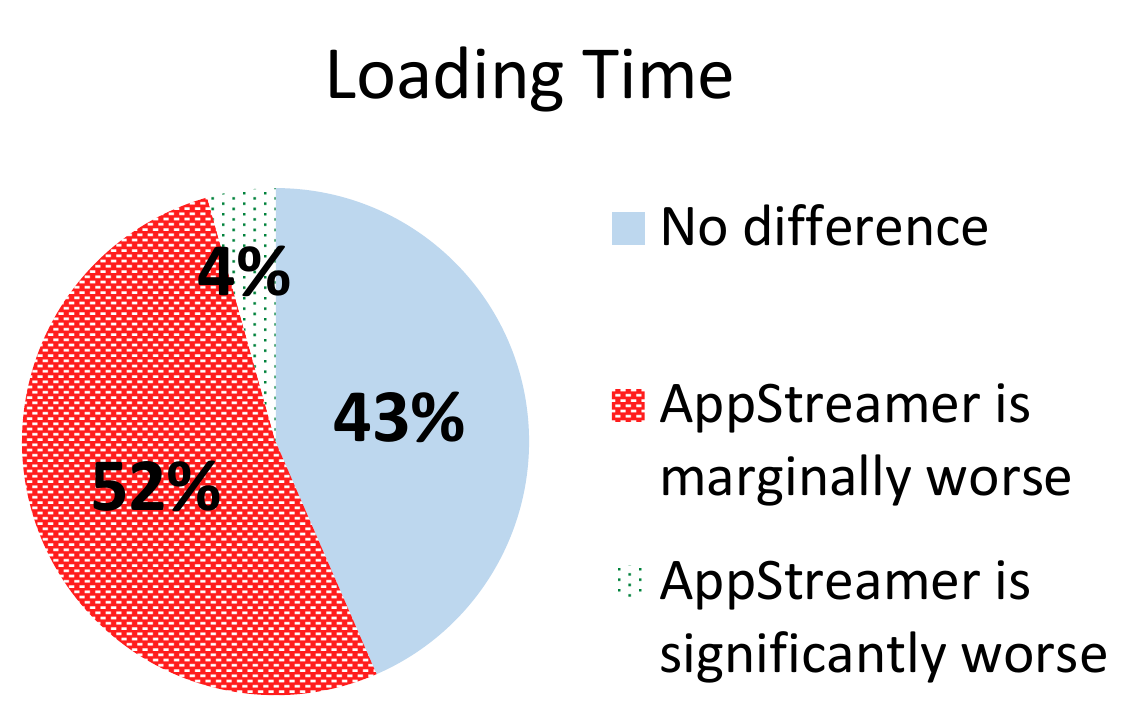}
\includegraphics[clip,trim=8cm 10.4cm 2cm 10.2cm,width=0.5\columnwidth]{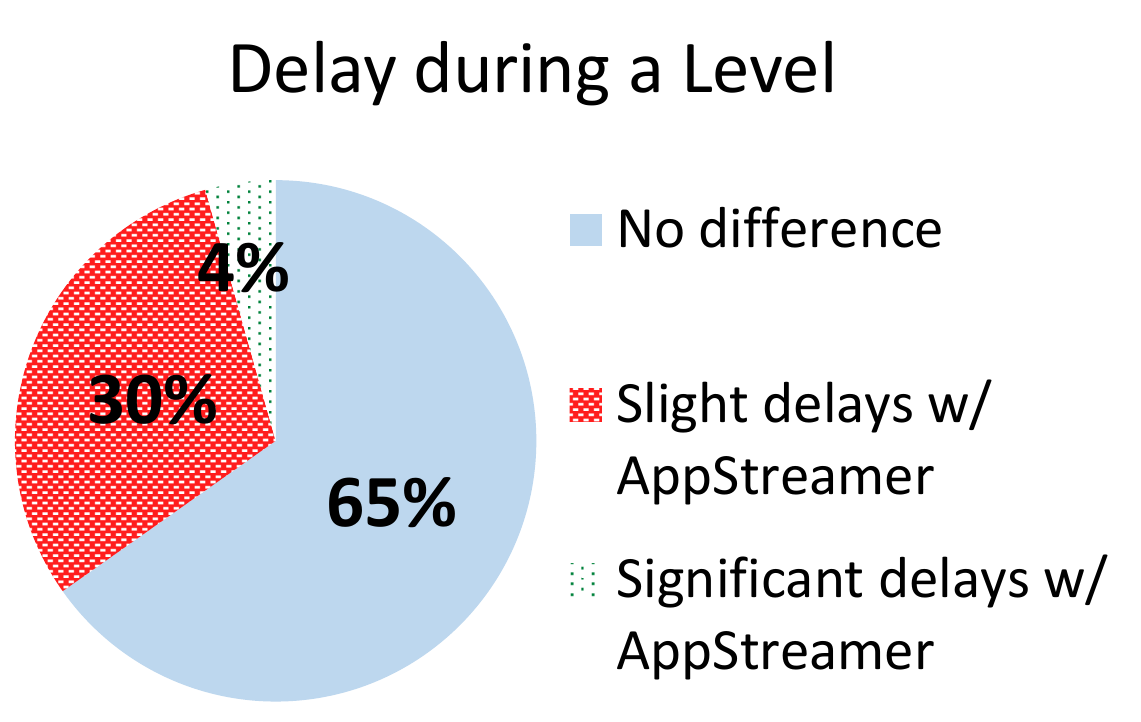}
\caption{User study results for Dead Effect 2 with 23 participants}
\label{fig:userstudy1}
\end{figure*}

The summarized responses to each question on the questionnaire are shown in Figure~\ref{fig:userstudy1}. 70\% of the participants rate the overall user experience of playing with \toolname the same as playing on an unmodified phone. The remaining 30\% rate the run with \toolname as marginally worse than baseline. There were no disruptions other than pauses, such as, crashes, hangs, or visual glitches, during gameplay with our technique in place.

On average, there were 336.2 KB of ``cache miss", blocks that the game tries to read but are not present on the phone, during each run with \toolname. This translates to 0.15 second of delay, for 28 minutes of gameplay, giving a 0.009\% delay. The cache hit rate is 99.87\%. The run that has highest amount of cache misses is affected by 1.52 seconds of delay. Compared to each level's loading time of roughly 20 seconds, this extra delay is barely noticeable. This shows that \toolname is able to predict and cache most of the necessary blocks before they are accessed for Dead Effect 2.

\subsection{User Study - Fire Emblem Heroes}
The second user study is done using Fire Emblem Heroes, with 26 users participating in the study. The study's setup is the same as the first user study with Dead Effect 2, except that the user is free to play any level for 20 minutes on each phone. The game has several different modes which are unlocked as the first few levels are completed, and some of these modes have multiple levels which the player can choose to play. Before playing, the participants are informed of these modes, and are instructed that they can switch to any mode and play any level as desired.

\begin{figure}[tb]
\centering
\includegraphics[clip,trim=2cm 9cm 1.7cm 9cm,width=0.8\columnwidth]{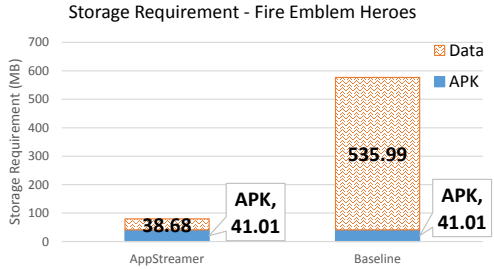}
\caption{Storage requirements for Fire Emblem Heroes}
\label{fig:storage2}
\end{figure}

The amount of storage used by Fire Emblem Heroes is shown in Figure~\ref{fig:storage2}. Baseline refers to the current state of practice which is storing the whole game on the phone. The amount shown for \toolname corresponds to the permanent storage used by files that are always stored on the phone. As the user plays the game, more blocks are downloaded on the fly. These blocks are stored in the temporary space and not shown in the figure. Overall, \toolname uses 79.69 MB of permanent storage, compared to 577 MB for baseline. This represents a 86\% saving of storage space.

\begin{figure*}[tb]
\centering
\includegraphics[clip,trim=8cm 10.2cm 4cm 10.2cm,width=0.4\columnwidth]{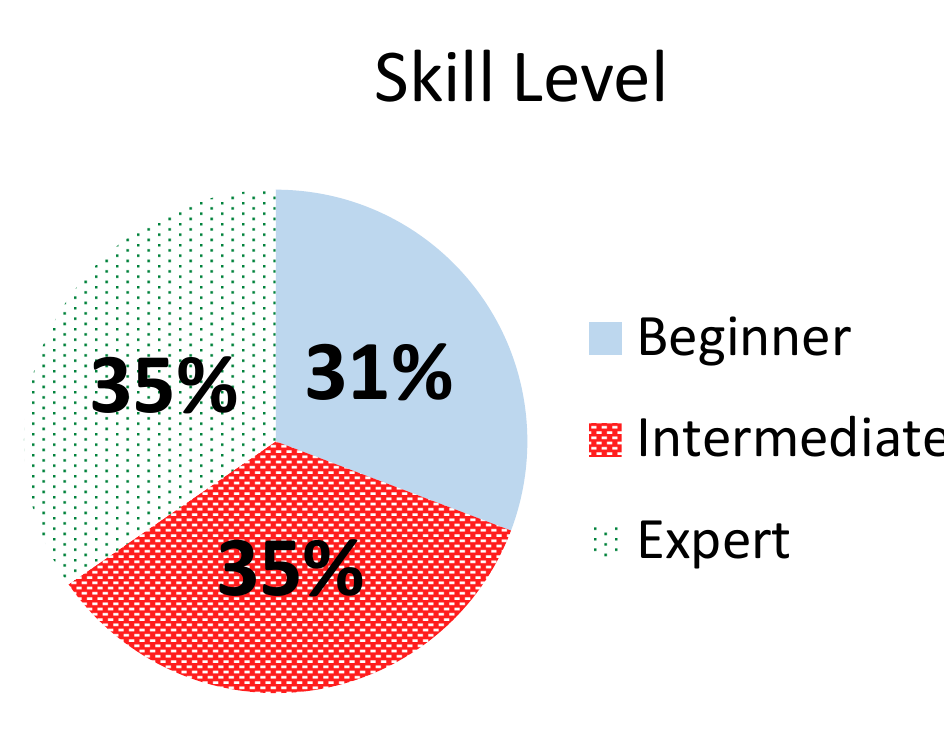}
\includegraphics[clip,trim=8cm 10.4cm 2cm 10.2cm,width=0.5\columnwidth]{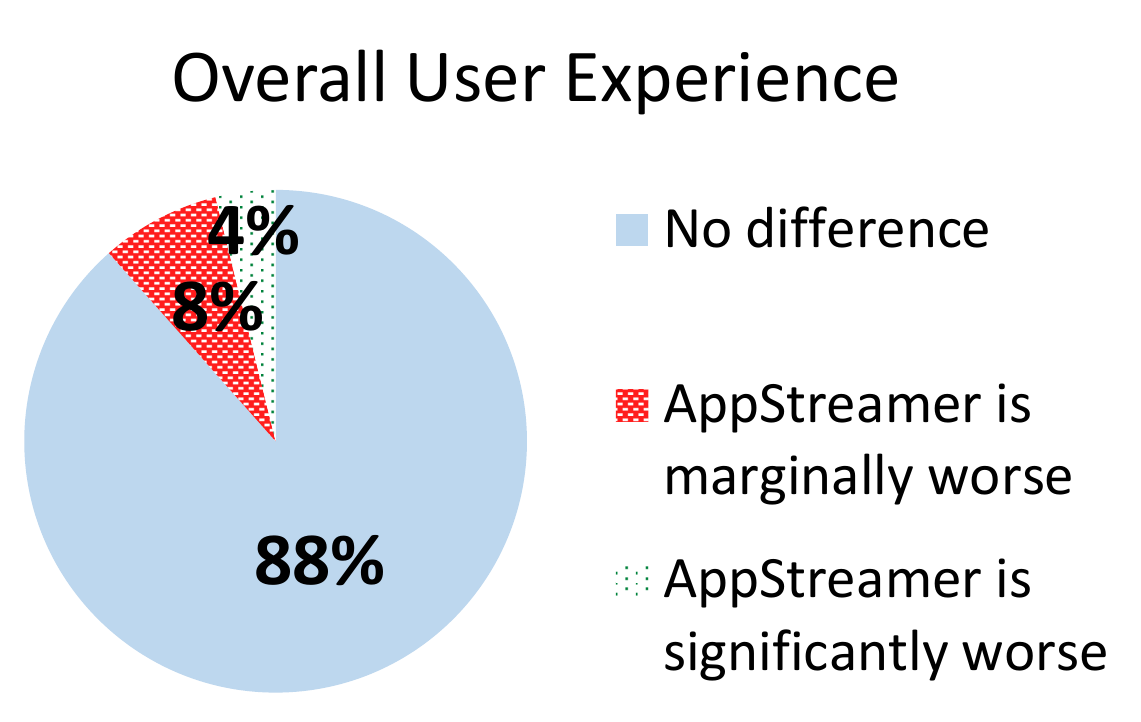}
\includegraphics[clip,trim=8cm 10.4cm 2cm 10.2cm,width=0.5\columnwidth]{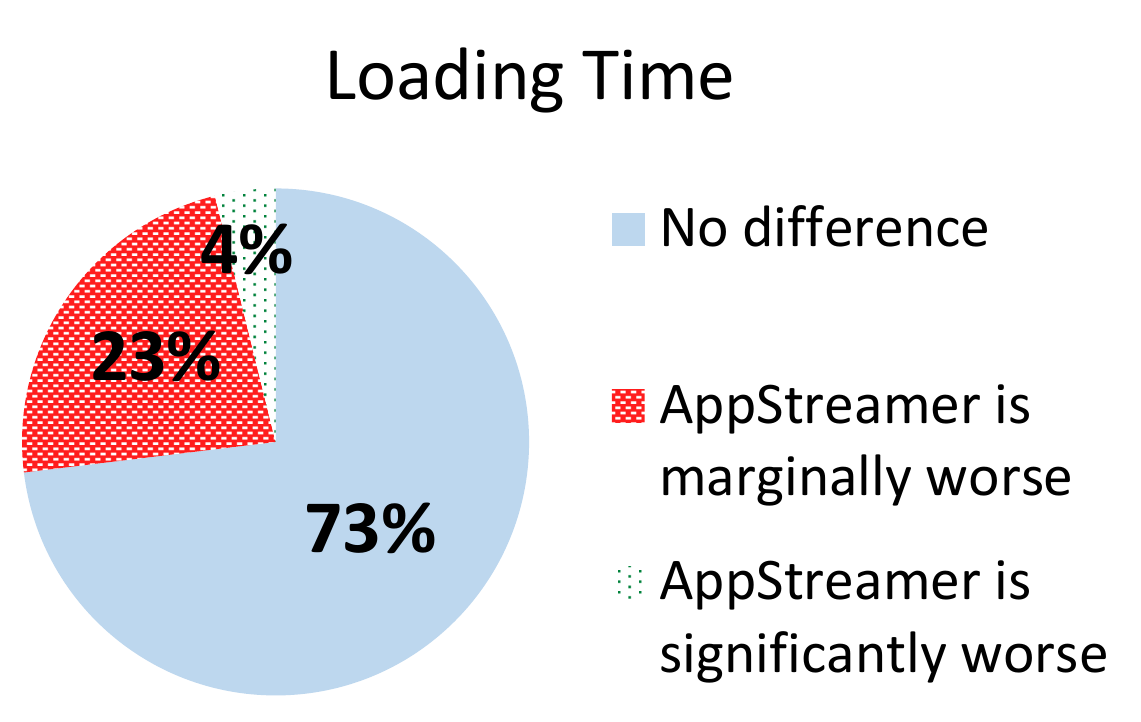}
\includegraphics[clip,trim=8cm 10.4cm 2cm 10.2cm,width=0.5\columnwidth]{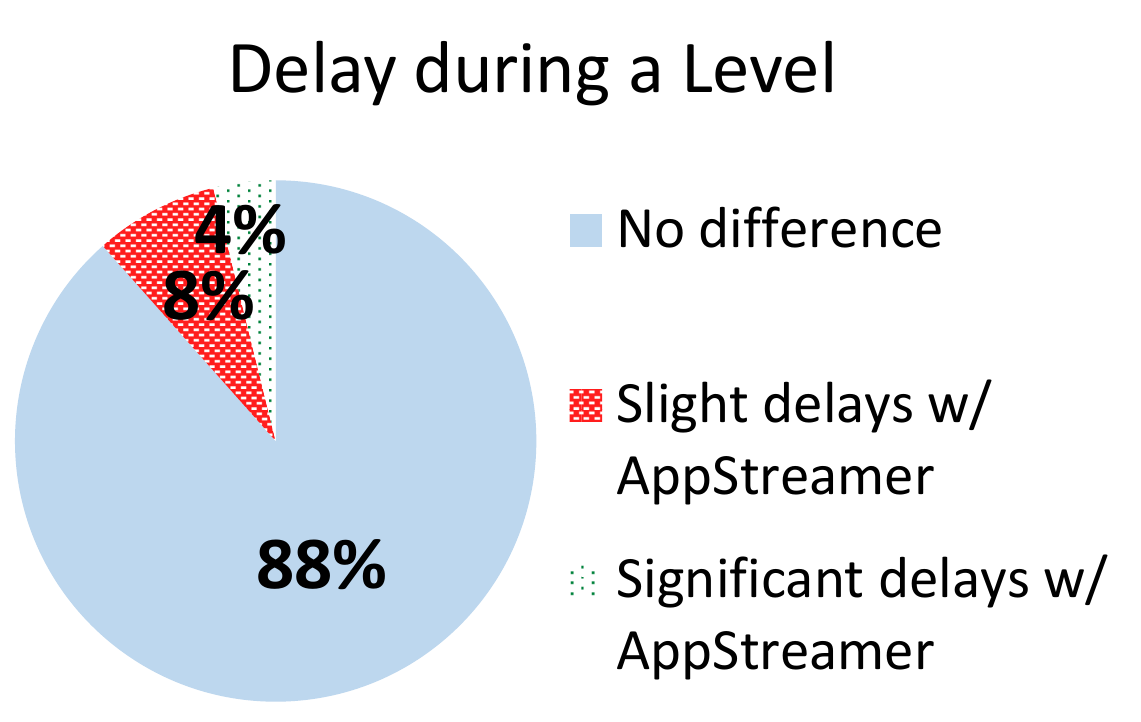}
\caption{User study results for Fire Emblem Heroes with 26 participants}
\label{fig:userstudy2}
\end{figure*}

The summarized responses to each question on the questionnaire is shown in Figure~\ref{fig:userstudy2}. 88\% of the participants rate the overall user experience of playing with \toolname the same as playing on an unmodified phone. Again, there were no disruptions other than longer loading time and delays during gameplay. Interestingly, two users comment that the run with \toolname was actually smoother. This might be explained by the fact that the delays before and after each level are dominated by network communication with the game server, rather than file reads, and the delay may depend on the current load of the game server.

On average, there were 1.63 MB of cache miss during each run with \toolname, which translates to 0.75 second of delay in 20 minutes of gameplay, giving a 0.0625\% delay. The cache hit rate is 97.65\%. The run that has highest amount of cache misses is affected by 5.12 seconds of delay. Nevertheless, the user still rated the overall user experience as no difference from the unmodified version. One user rates the run with \toolname as significantly worse due to significant delays. However, the communication log on the cloud storage server indicates that only 0.97 MB of blocks were missed and needed to be fetched urgently. This translates to 0.45 second of delay, which should be barely noticeable. Overall, the results of this user study show that \toolname is able to predict and cache most of the necessary blocks before they are accessed, even when there are different branches for different users in the gameplay.

\subsection{Comparison with Prior Work}
\label{sec:prior-comparison}

Here we compare the bandwidth consumption and latency of \toolname to state-of-the art cloud gaming systems.  One challenge in thin client gaming is that users are disturbed by latencies higher than 60 ms \cite{quax2004objective}. Lee {\em et al.} addresses this problem by using speculative execution, which can mask up to 128ms of latency, at the cost of using between 1.51 and 4.54 times as much bandwidth as standard cloud gaming \cite{outatime}.  Using speculative execution requires access to and modification of the source code of the game, so we could not directly compare \toolname to speculative execution.  However, we tested the performance of a thin client, GamingAnywhere, an open source cloud gaming system \cite{huang2014gaminganywhere}.

\begin{figure}[h]
\centering
\includegraphics[clip,trim=5.4cm 10.2cm 6.2cm 10.7cm,width=0.49\linewidth]{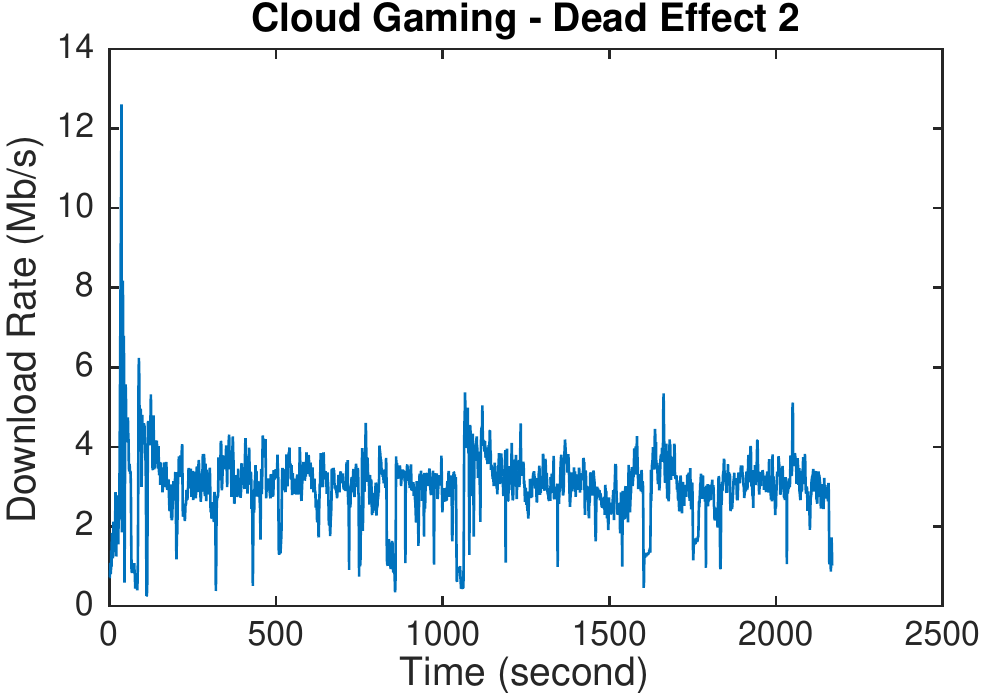}
\includegraphics[clip,trim=5.4cm 10.2cm 6.2cm 10.7cm,width=0.49\linewidth]{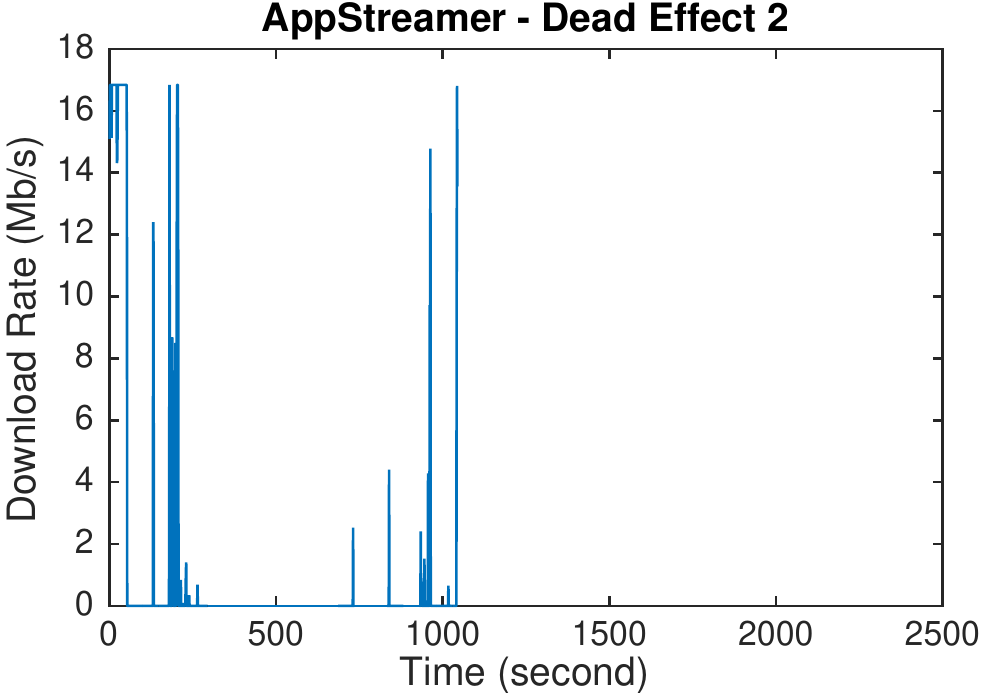}
\includegraphics[clip,trim=5.4cm 10.2cm 6.2cm 10.1cm,width=0.49\linewidth]{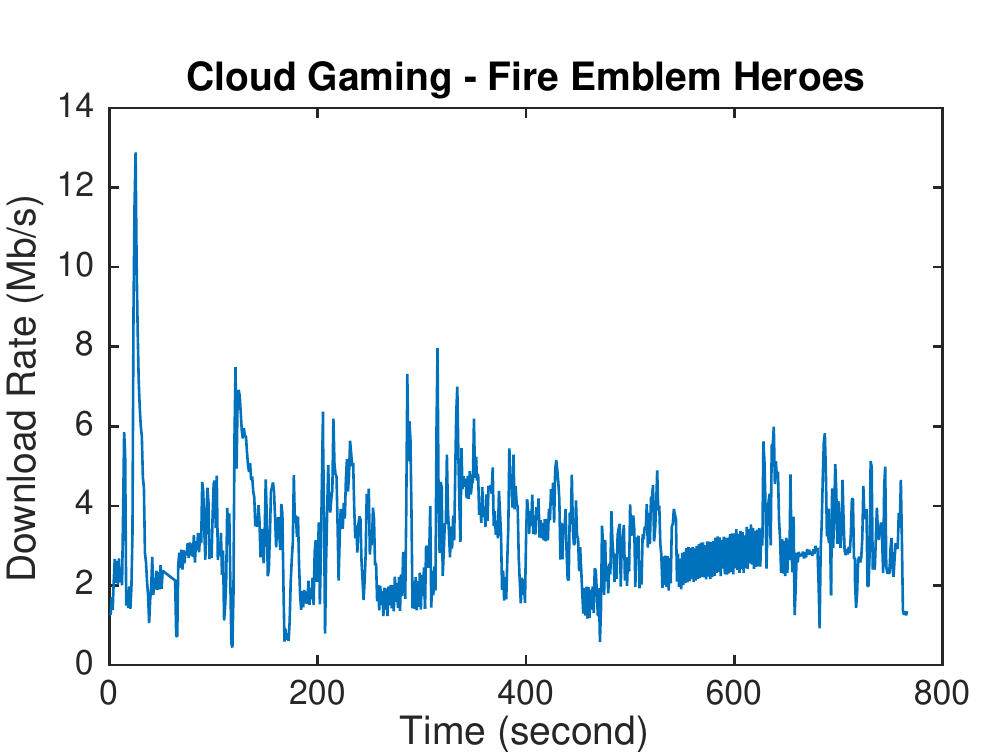}
\includegraphics[clip,trim=5.4cm 10.2cm 6.2cm 10.1cm,width=0.49\linewidth]{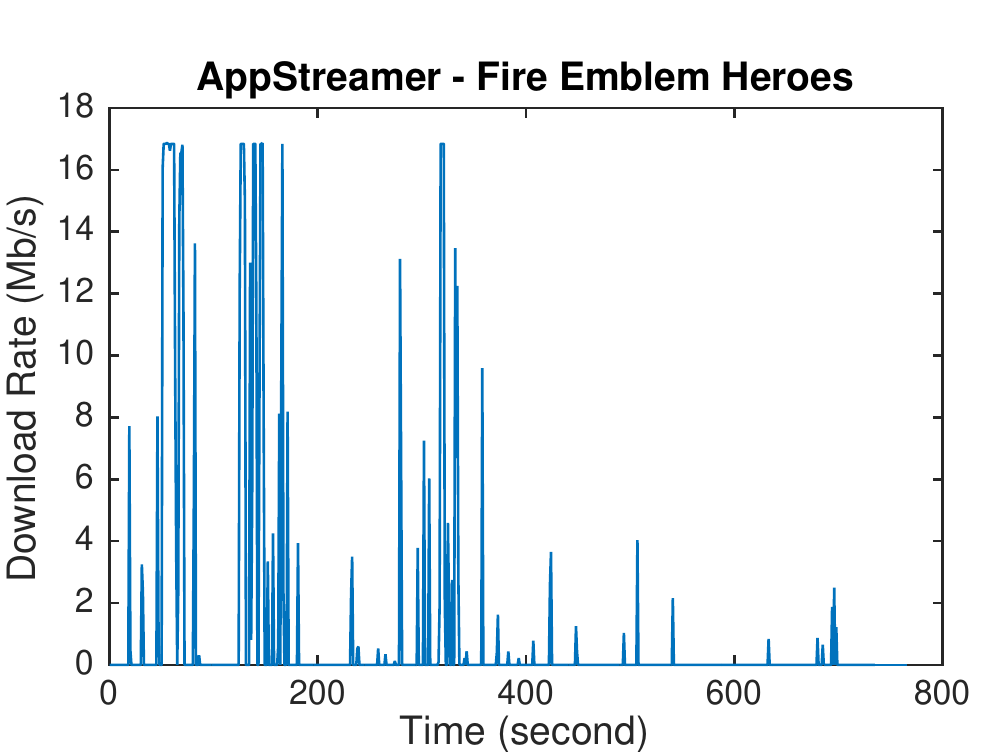}
\caption{Comparison of bandwidth consumption between cloud gaming (left) and \toolname (right) for both games}
\vspace{-0.3cm}
\label{fig:GamingAnywhere}
\end{figure}

\noindent {\bf Cloud gaming} \\
In order to determine the bandwidth usage of a thin client model, we ran Nox, an Android emulator, and GamingAnywhere on a server and the GamingAnywhere client on a smartphone. We tested both Dead Effect 2 and Fire Emblem Heroes, and recorded the download bandwidth usage of the GamingAnywhere application on the smartphone. Data uploaded from the smartphone consists of encoded input events (such as swipes and taps), and data downloaded consists of audio and video of the game being streamed. The bandwidth usage of cloud gaming and \toolname are shown in Figure~\ref{fig:GamingAnywhere}. We found that for Dead Effect 2, cloud gaming uses 3.01 Mb/s while \toolname only uses 706 Kb/s on average. For Fire Emblem Heroes, cloud gaming uses 3.20 Mb/s while \toolname only uses 745 Kb/s on average. This shows that traditional cloud gaming is a lot more bandwidth intensive than our file block streaming approach, with 4.3X higher bandwidth requirement for our two target games.

Compared to the baseline where the entire application is downloaded before it can be used, \toolname likely uses more bandwidth through the more costly cellular connection. This can be alleviated in two ways. First, as long as the necessary blocks are never evicted after they are downloaded, the total bandwidth usage cannot exceed the total size of the application. Even when space runs out, LRU eviction policy helps prioritize keeping blocks that are more likely to be accessed. Second, \toolname can be extended so that block prefetching is done more aggressively when the device is on a Wi-Fi connection, so that less fetching is needed when the device is on a cellular connection.

As mentioned earlier, latency is a key measure of usability of cloud gaming tools.  Latency can be very visible and annoying to users, as the time between every user input (\eg, a screen tap) and the frame that shows the effect of that input is the network round-trip time, plus video/audio encoding and decoding delay. The network round-trip time depends largely on the distance between the cloud gaming server and the client. Based on typical placement of gaming servers, a typical round-trip time is 100 ms~\cite{sundaresan2011broadband}. On the other hand, \toolname is not as heavily affected by latency as much as cloud gaming approaches, since speculative block fetches are mostly in batches and can be easily pipelined. The urgent block fetches are affected by the latency, but the amount of urgent block fetches is typically small. As a back-of-the-envelope calculation, for Dead Effect 2, on average 84.05 out of 64,858 blocks accessed are fetched urgently. Fetching each block requires 100 ms + 4 KB / 17.4 Mbps = 101.8 ms. Thus, the overall delay is $\frac{84.05}{64858}\times 101.8 = 0.13$ ms, which is much smaller than the constant 100 ms in the cloud gaming approach. For Fire Emblem Heroes, \toolname's overall delay is $\frac{416.96}{17731}\times 101.8 = 2.39$ ms.


\noindent {\bf Block-wise predictor} \\
In addition to the cloud gaming approach, we also compare \toolname to a simple file access prediction algorithm that operates at the block granularity, which we call \emph{BlockPairLookup}. In the training phase, it stores all pairs of blocks $(B_i, B_j)$ such that $B_j$ is read within the lookahead time $L$ after $B_i$ is read. In the online phase, when a block $B_i$ is accessed, it predicts all $B_j$'s where $(B_i, B_j)$ is in its memory.

We run the BlockPairLookup algorithm for both games and compute the delay and amount of unnecessary blocks downloaded using our simulator. We find that it has excessive memory utilization---20.5 GB with 30 seconds lookahead with Dead Effect 2 and 4.6 GB with 60 seconds lookahead with Fire Emblem Heroes. Both would be infeasible on today's mobile devices. 
For Dead Effect 2, with BlockPairLookup, average delay per run is 6.32 seconds (8.4X of \toolname), and 74.39 MB of unnecessary blocks are downloaded (1.1X of \toolname). For Fire Emblem Heroes, average delay per run is 7.32 seconds (18.8X), and 64.10 MB of unnecessary blocks downloaded (1.1X). Because BlockPairLookup's predictions are always a superset of \toolname's predictions, the higher delay is likely due to the unnecessary blocks that are put in the download queue delaying the download of necessary blocks and the inefficiency of requesting a single block at a time. 
This shows that models that operate on single block granularity incur too much memory and delay and are thus impractical.

\subsection{Microbenchmarks}
\label{sec:microbenchmarks}
In this section, we evaluate how different parameters affect the results. The parameters studied are $\delta$, $\tau$, $p_{stop}$, $L$, $minSuperblockSize$, $p_{download}$, and $B_{initial}$, described in Section~\ref{sec:design}, as well as buffer size and network connection speed. The results are generated based on trace-based simulation. In the simulation, first training data is used to train a Markov model. Then, file reads from the test data is replayed and given as input to the Markov model. Blocks predicted by the model that are not already present on the phone are logically fetched from the storage server, with network speed fixed to a certain value to simulate real network conditions. In the case where buffer size is limited, we employ the LRU policy to evict blocks from the limited storage available.

Since there are many parameters, we conduct the microbenchmarks by varying one parameter at a time, and fixing the rest of the parameters to the optimal value. Optimal values are chosen by carefully weighing the tradeoff between delay and false positives, with higher weight given to delay, as it has a direct impact on the user experience. The values are $\delta$ = 0.1 second, $\tau$ = 0.9, $minSuperblockSize$ = 17, $B_{initial}$ = 122 MB (excluding APK), $p_{stop}$ = 0.01, $L$ = 60 seconds, and connection speed = 17.4 Mbps. By default, we do not set a limit on temporary storage used to store fetched blocks. The average length of each run is 1,653 seconds. Due to limited space and the fact that the results show the same trends, we omit the microbenchmark results for Fire Emblem Heroes, and show only results for Dead Effect 2.
The output metrics are delay and false positives, defined as predicted blocks that are not read by the game within 8 minutes of being predicted. Delays that are long or frequent enough can ruin the user experience, while false positives incur extra network bandwidth and energy cost.

\begin{figure*}[tb]
\centering
\includegraphics[clip,trim=5cm 10cm 5.3cm 10.4cm,width=0.41\columnwidth]{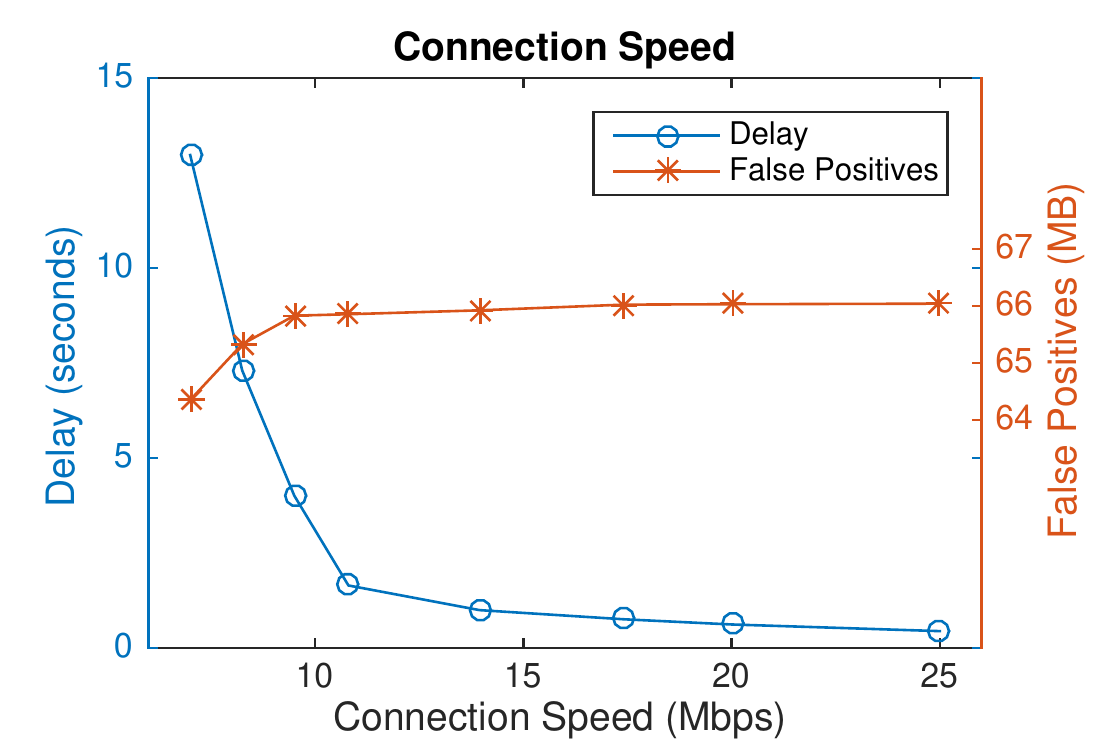}
\includegraphics[clip,trim=5cm 10cm 5.3cm 10.4cm,width=0.41\columnwidth]{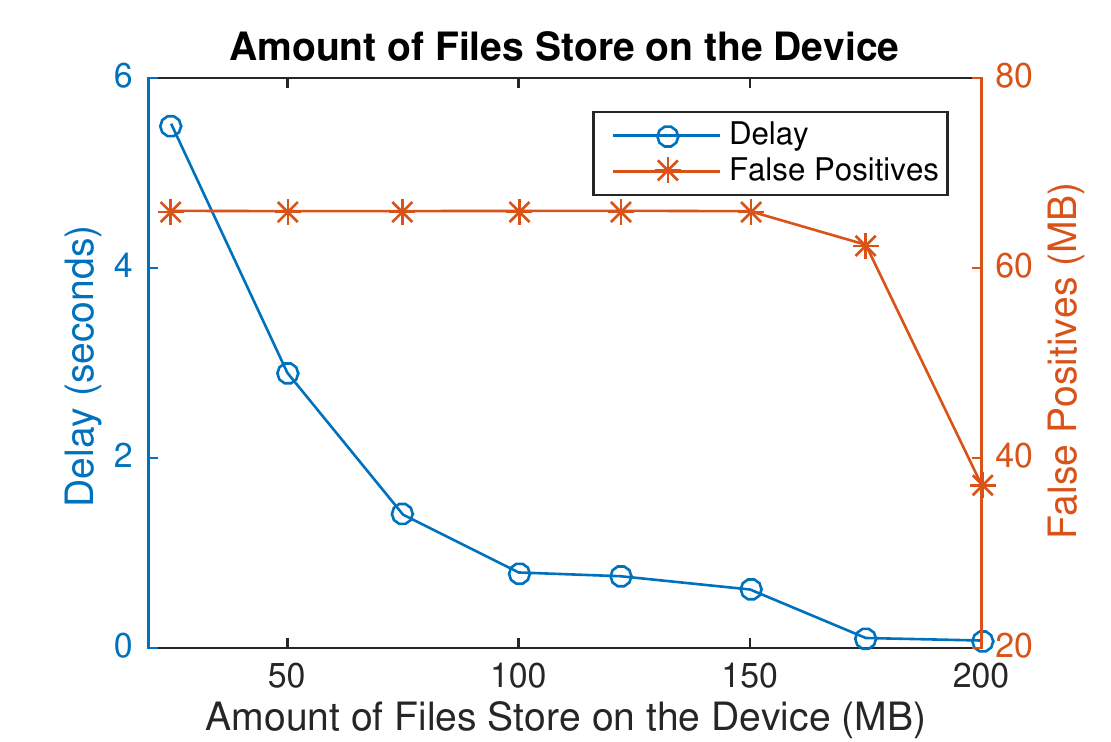}
\includegraphics[clip,trim=5cm 10cm 5.3cm 10.4cm,width=0.41\columnwidth]{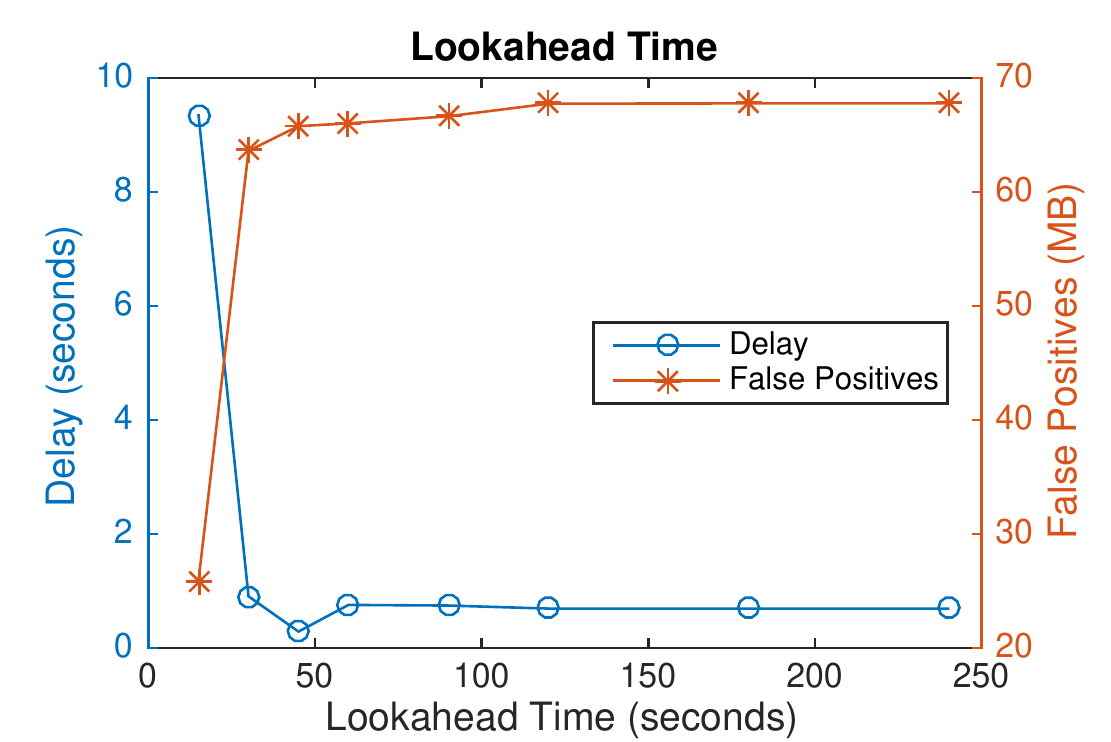}
\includegraphics[clip,trim=5cm 10cm 5.3cm 10.4cm,width=0.41\columnwidth]{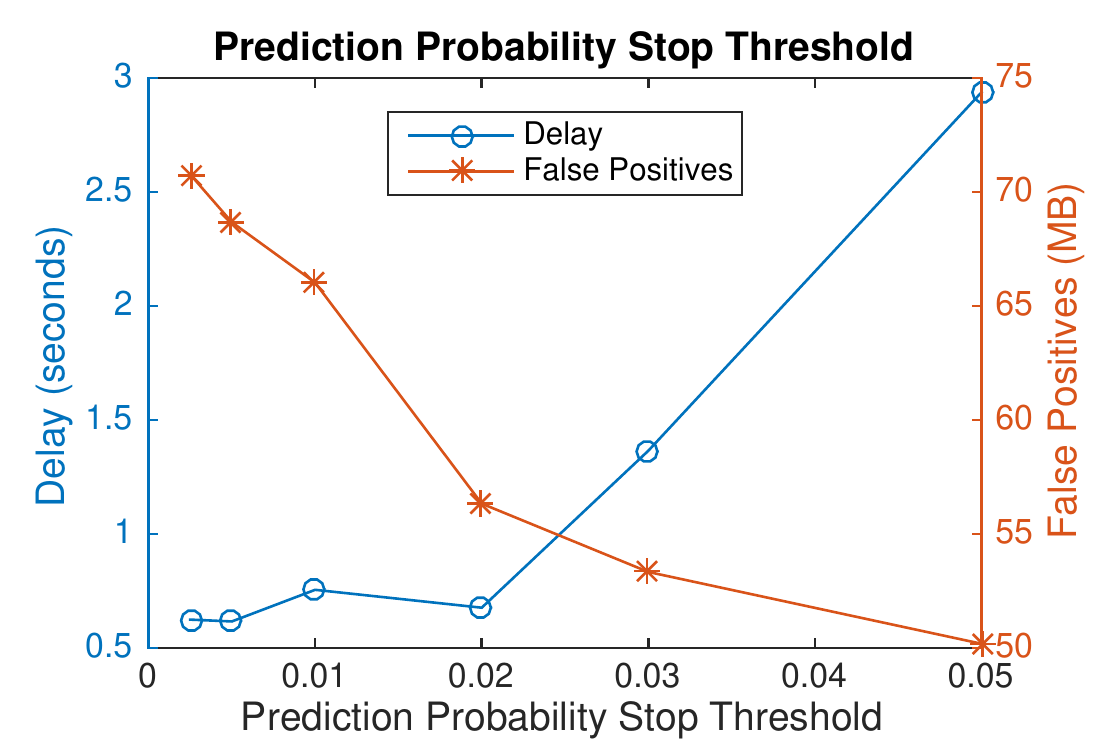}
\includegraphics[clip,trim=5cm 10cm 5.3cm 10.4cm,width=0.41\columnwidth]{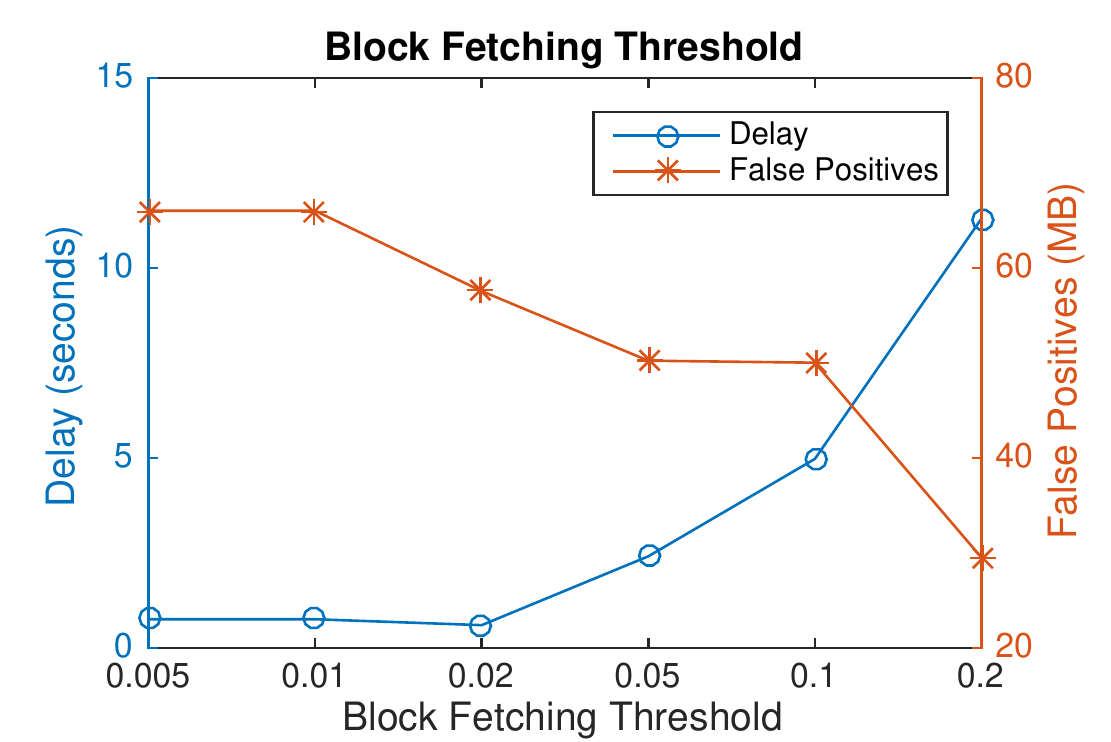}
\includegraphics[clip,trim=5cm 10cm 5.3cm 10.4cm,width=0.41\columnwidth]{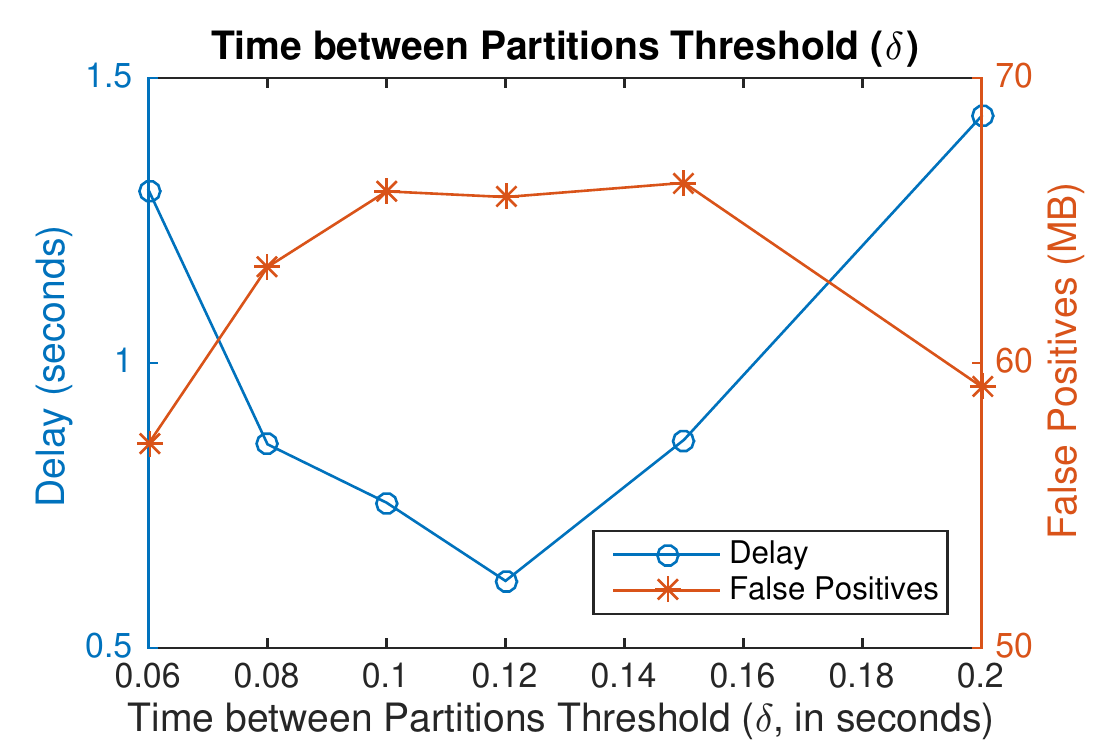}
\includegraphics[clip,trim=5cm 10cm 5.3cm 10.4cm,width=0.41\columnwidth]{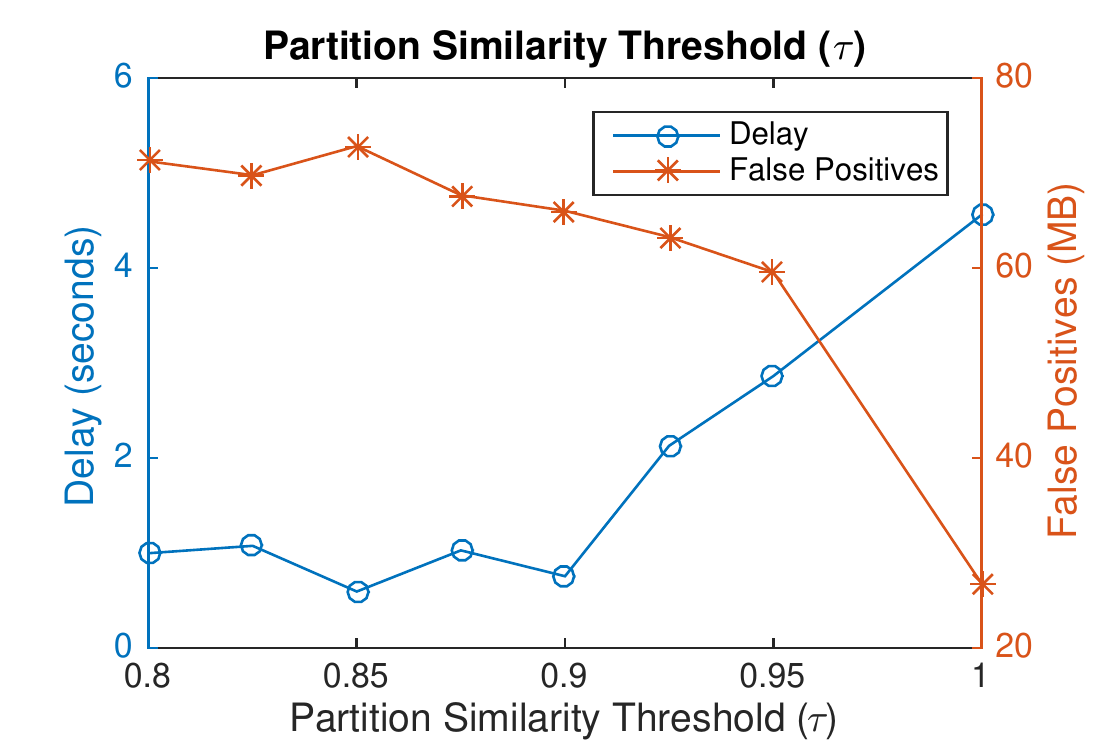}
\includegraphics[clip,trim=5cm 10cm 5.3cm 10.4cm,width=0.41\columnwidth]{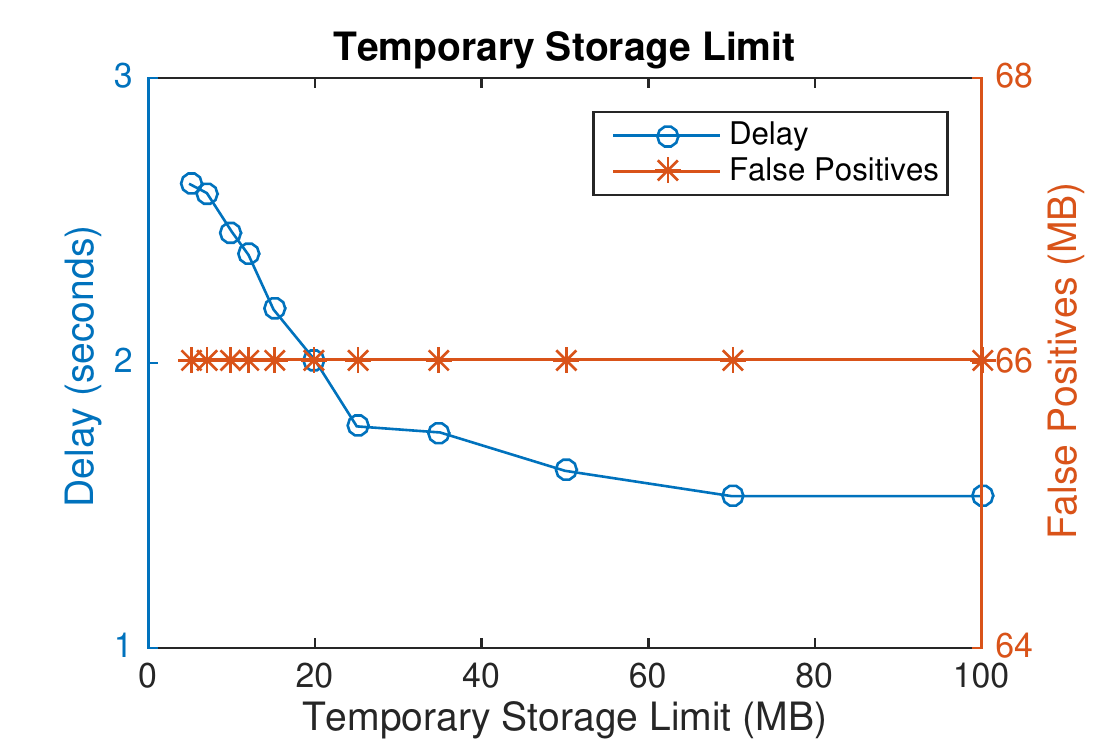}
\includegraphics[clip,trim=5cm 10cm 5.3cm 10.4cm,width=0.41\columnwidth]{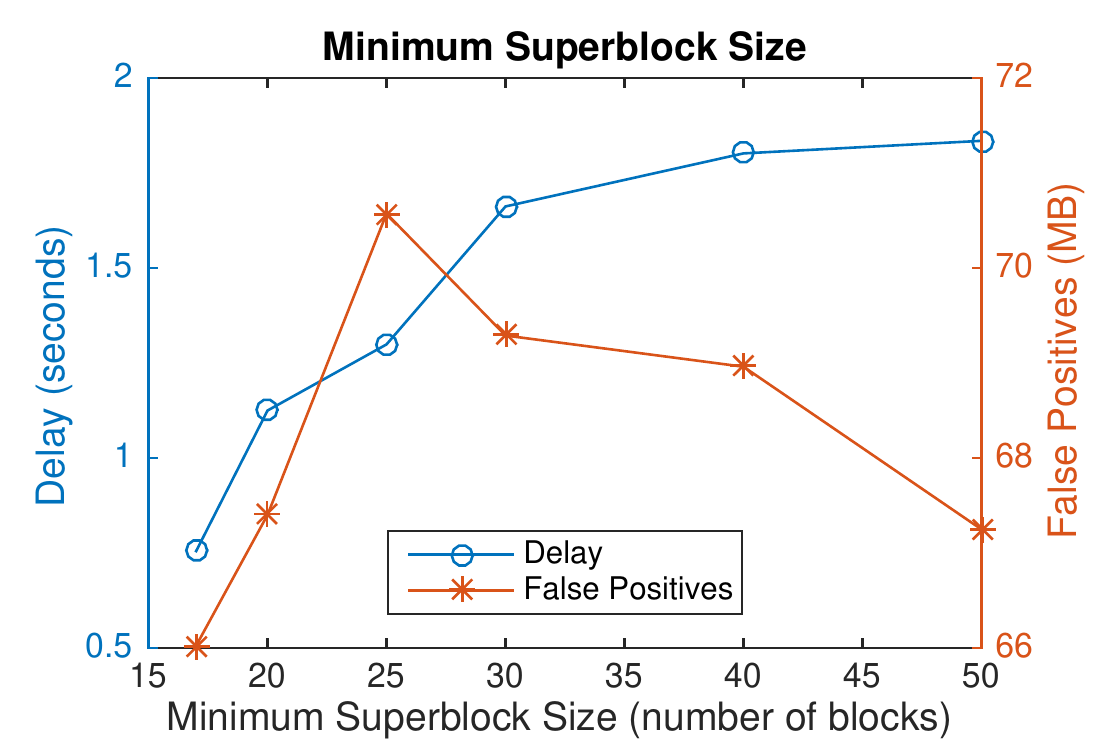}
\caption{Microbenchmarks for Dead Effect 2}
\label{fig:micro1}
\end{figure*}

The results are shown in Figure~\ref{fig:micro1}. First, with the optimal 
parameter values, the amount of false positives is 66 MB. However, if the 
playing session were longer, the amount of false positives will not 
necessarily increase proportionately, because there is a limit to how much 
data is downloaded, namely the total size of the application. 

Now, we look at how each parameter affects the results. In addition to average worldwide LTE speed of 17.4 Mbps, we also include average U.S. LTE speed of 13.95 Mbps and average worldwide WiFi speed of 10.8 Mbps. As expected, higher connection speed leads to lower delay. Even the lower WiFi speed of 10.8 Mbps is enough to keep the delay small, but speed lower than that will result in large delay. Connection speed has a negligible effect on the false positives.
Next, we look at the amount of initial files cached on the phone, denoted $B_{initial}$. Higher value gives lower delay and false positives, at the cost of higher storage requirements. Delays are virtually eliminated at $B_{initial} \geq$ 175 MB. This represents a storage savings of 84\%. At the higher end of 200 MB, the amount of false positives is also reduced.

Recall that when making predictions, our Markov model relies on two stopping criteria to keep the computation tractable: lookahead time, denoted by $L$, and probability stop threshold, denoted by $p_{stop}$. From the results, as long as the lookahead time is at least 30 seconds, the delay remains constantly low and the amount false positives is largely constant. When the lookahead time is too low, delay increases significantly. Probability stop threshold is somewhat similar. As long as the value is 0.02 or lower, delay remains relatively constant. Higher value leads to higher delay. The amount of false positives is lower when $p_{stop}$ is higher, as early stop means fewer blocks get predicted.
The block fetching threshold, denoted by $p_{download}$, affects the final decision of whether or not to download blocks in the predicted merged cluster, based on predicted probability. It directly influences the amount of false positives, with higher threshold resulting in lower false positives. However, the delays are kept at an acceptable level only when $p_{download}$ is 0.02 or lower.

Time between partitions threshold, denoted $\delta$, controls how consecutive blocks are merged into the same partition. Lower value leads to more partitions that are smaller. The results clearly show that 0.12 is the optimal value with respect to delay. 
This amount represents the upper limit of the amount of computation (\eg, image decoding) the application does between chunks of data in the same batch of read. 
Partition similarity threshold, denoted $\tau$, controls merging of two similar partitions within the same trace. A value of 1 means the two clusters need to contain the exact same blocks in order to be merged. The results show that values between 0.8 and 0.9 produce similarly low delay, while higher values result in higher delay.

Temporary storage limit sets a hard storage limit for storing blocks fetched speculatively. This does not include the APK and files that are always stored on the phone. In reality, this buffer can be shared by all applications as long as they do not run at the same time.
The results show that a small 75 MB buffer is already as good as an infinitely large buffer. Thus, the amount of temporary space required by \toolname is very small. 

The minimium superblock size 
serves as the stopping criterion of the first step of the process of generating superblocks. Lower value leads to more precise model and predictions, but incur longer training time. The results confirm that lower values are always better than higher values in terms of delay. However, we could not complete the benchmark using values lower than 17, as the training time suddenly jumps from a few minutes to several hours.

%% file: sec_discussion.tex
\section{Discussion}
\label{sec:discussion}
{\textbf{Cloud Gaming.}
Cloud gaming is an alternative to AppStreamer, since the game runs on the cloud and only the video and audio are streamed to the mobile device. As mentioned in Section~\ref{sec:prior-comparison}, the main drawbacks of cloud gaming are high latency and high bandwidth usage. Within the foreseeable future, the amount of bandwidth cloud gaming uses makes it prohibitively expensive for most users. Latency can be reduced by moving the cloud closer to the users, but occasional packet losses (common in wireless communication) still degrades user experience.
}



Our solution performs all computation on the mobile device, relying on the cloud storage server only for storage. Cloud gaming, on the other hand, performs all computation on the server, at the expense of bandwidth usage and latency. There is likely a middle ground whereby computation can be divided between the device and the cloud server. This can benefit from the long line of prior work in computation offloading from resource-constrained devices \cite{gordon2012comet, shi2014cosmos, kao2017hermes} but will have to be repurposed due to some domain-specific requirements, including stringent low latency and high degree of dynamism in gameplay.


{\textbf{Trace Collection and Model Training.}
Although \toolname will work for all mobile applications without any modification to the applications themselves, it is necessary to collect file access traces from a large number of users to cover most file access patterns. The trace collection system can be built into the operating system itself, and the user can decide whether to contribute to trace collection. For the first version of the application or when the application is updated, the developer may decide to ship it without \toolname, or collect the traces privately (e.g., from testers) beforehand. Model training is done by the developer and takes place offline. Parameter optimization can easily be parallelized. We envision that the developer is responsible for trace collection and model training because they should be in control of the user experience of their own application, and the computation is distributed, compared to relying on a single entity.}

{\textbf{Technological Advancements.}
\toolname is still relevant even in the face of constantly improving storage capacity on smartphones. As more storage becomes available on smartphones, developers also take advantage of it more resulting in a larger size of the games. With higher screen resolution, artists tend to use more detailed texture and 3D models to improve their games’ visuals.}


{In addition to reducing storage requirements, \toolname also helps reduce application installation time, since only a small part of the application needs to be downloaded before the user can use it, and the rest is downloaded only as needed, just like video streaming.}

%% file: sec_conclusion.tex
\section{Conclusion}
\label{sec:conclusion}

We set out to see how to reduce the storage pressure caused by resource-hungry applications on mobile devices. We found that mobile games were a significant contributor to the problem. We had the insight that mobile games do not need all the resources all the time. So if it were possible to predict which resources would be needed with enough of a lookahead, then they can be prefetched from a cloud storage server and cached at the device and thus not cause any stall for the user. We achieve this goal through the design of \toolname, which uses a Markov Chain to predict which file blocks will be needed in the near future and parametrizes it such that the model can be personalized to different speeds and gameplay styles. We show that for two popular third-party games, \toolname reduces the storage requirement significantly (more than 85\%) without significantly impacting the end user experience. This approach can also help to reduce the startup delay when an app is being downloaded and installed as well as to reduce stalls with cloud gaming by pre-fetching the required resources. 

\section{Acknowledgments}
\label{sec:acknowledgments}
This material is based in part upon work supported by the National Science Foundation under Grant Numbers CNS-1513197 and CNS-1527262 and gift funding from AT\&T. Any opinions, findings, and conclusions or recommendations expressed in this material are those of the authors and do not necessarily reflect the views of the sponsors.